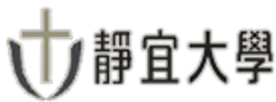

周憲辛 <hhchou@gm.pu.edu.tw>

# [化學季刊] 稿件接受刊登通知_chem.202512_83(4).008
2 封郵件

---

**web cslt** <webpmcslt@gmail.com>　　　　　　　　　　　　　　　　　　2025年12月29日上午10:26
收件者: hhchou@pu.edu.tw

周憲辛副教授 您好：

很高興通知您論文已審查完畢，並接受刊登在 12 月號的【化學季刊】中。編輯部將檢查後進行排版，後續將通知您進行校稿。

稿件編號：chem.202512_83(4).008
論文名稱：全光譜吸收材料：分子設計與光伏應用挑戰

「　周教授綜整回顧文獻成果，可以清楚看出，全光譜吸收材料在太陽能光伏應用中所扮演的角色，已逐漸由早期單純追求吸收範圍的擴展，轉向更為細緻且系統性的效率權衡問題。既有文獻普遍指出，雖然吸收光譜的展寬有助於提升入射光子的利用比例，但在實際元件中，光收成效率的提升往往同時伴隨能隙縮小、能階下移與電荷轉移驅動力降低等現象，使得電子注入、電荷分離或再生動力學受到限制。此一結構–性質–效能之間的連動關係，已成為全光譜吸收材料設計中無法迴避的核心議題。
　　從電子結構觀點而言，文獻一致顯示，不同波段吸收對應之電子躍遷型態本質上存在差異。紫外與可見光區多由 □→□*或部分 n→□*躍遷主導，而近紅外吸收則往往涉及高度電荷轉移特徵之低能量激發態。為了在單一分子系統中同時涵蓋上述多種躍遷，研究者必須透過降低鍵長交替、延伸π共軛、導入推–拉型結構或穩定醌式共振等方式來調控前沿分子軌域。然而，相關研究亦反覆指出，過度壓低能隙或強化分子內電荷轉移，可能導致躍遷偶極矩下降、非輻射鬆弛速率上升，甚至影響材料的化學與光穩定性。由此可見，寬光譜吸收本身並非獨立可優化的目標，而是與多項光物理與動力學參數高度耦合。
在光伏元件層面，不論是染料敏化太陽能電池或有機太陽能電池，既有研究皆指出短路電流密度的提升必須建立在有效電子注入、激子解離與電荷收集等程序皆能順利進行的前提之下。多項文獻案例顯示，當全光譜吸收材料之 LUMO 能階過低或與半導體導帶匹配不佳時，即便吸收範圍顯著拓展，實際 IPCE 或整體光電轉換效率仍可能受限。此類結果進一步凸顯，能階配置與界面動力學在全光譜材料設計中具有與吸收光譜同等重要的地位。
在分子工程策略方面，本文所整理之研究顯示，各類設計手法皆各具優勢與限制。推–拉型分子與 D–A 交替共軛系統有助於調控能階與引入分子內電荷轉移特徵，但其效能高度仰賴推拉強度與結構平衡；醌式共振與多亞甲基結構可有效縮小能隙並推進近紅外吸收，卻常伴隨可見光區吸收不連續或動力學驅動力不足的問題；多環芳香烴與融合型結構雖能提供高度離域化的電子結構與寬廣吸收，卻可能因過度平面化而引發聚集與傳輸限制。這些文獻結果共同指出，單一設計策略難以同時滿足光譜完整性、能階匹配與動力學效率三者的需求。
值得注意的是，筆者認為過去研究亦顯示，金屬錯合物系統透過 MLCT 或 LMCT 躍遷，能在不極端壓縮能隙的情況下提供寬廣吸收，並藉由重原子效應與自旋–軌域耦合穩定低能量激發態。然而，其吸收強度與穩定性仍高度依賴配位基設計與整體分子結構的精細調控。此類研究進一步說明，全光譜吸收的實現往往來自多重機制的協同，而非單一結構因素的極端化。
　　整體而言，過往文獻顯示，全光譜吸收材料的有效應用，已逐漸由「單一分子光譜最佳化」轉向「分子、半導體與界面系統之協同設計」。共敏化策略、多材料搭配與能階互補的概念，正是回應單一分子設計受限的實務解方。這些研究結果共同指向一個重要結論：在光伏應用中，真正值得追求的並非光譜覆蓋的極限，而是在吸收能力、能階配置與電荷轉移動力學之間取得整體效率的最適平衡。
　　本篇為綜整既有研究的評論性文章，本文成功彙整並串聯了分子光譜工程、電子結構設計與光伏元件物理之間的關鍵脈絡，清楚呈現全光譜吸收材料發展所面臨的共通挑戰與設計取捨。透過跨系統、跨材料類型的文獻對照，作者縱觀文獻結論為讀者提供了一個有助於重新思考「寬吸收與高效率關係」的整合性視角，對後續研究在材料設計策略與研究方向的選擇上，具有明確且實質的參考價值。」

---

化學季刊編輯部 張芷華 敬上
E-mail: webpmcslt@gmail.com
Tel: 02-5572-8574

---

**Hsien-Hsin Chou** <hhchou@gm.pu.edu.tw>　　　　　　　　　　　　　　2025年12月29日下午4:29

# Panchromatic Absorbing Materials: Molecular Design and Challenges in Photovoltaic Applications

Hsien-Hsin Chou*

Department of Applied Chemistry, Providence University, Taichung, Taiwan

**Abstract**

Panchromatic absorbing materials are widely regarded as a key strategy for enhancing solar energy utilization and photocurrent generation. However, in artificial molecular systems, broadening the absorption spectrum is often accompanied by fundamental challenges, including bandgap narrowing, poor energy-level alignment, and limited charge-transfer kinetics, indicating that pursuing broadband absorption alone is insufficient to guarantee high photovoltaic performance. This article examines the relationship between design strategies and performance of panchromatic absorbing materials from the perspectives of molecular engineering and photovoltaic devices, with particular emphasis on the delicate balance among molecular electronic structure, charge-transfer characteristics, interfacial energy-level alignment, as well as electron injection, regeneration efficiency, and energy losses. Ultimately, the molecular design of panchromatic photovoltaic materials should move beyond molecular-level optimization toward synergistic tuning among molecules, semiconductors, and electrolytes or active-layer materials, thereby providing concrete conceptual guidance for achieving "efficiency optimization" rather than simple "spectral maximization."

# 全光譜吸收材料：分子設計與光伏應用挑戰

周憲辛*

靜宜大學應用化學系

## 摘要

全光譜吸收材料被視為提升太陽能利用率與光電流的關鍵策略，然而在人工分子系統中，吸收展寬往往伴隨能隙縮小、能階匹配不佳與電荷傳遞動力學受限等根本性挑戰，顯示單純追求寬光譜吸收不足以保證高效率的光伏表現。本文從分子工程與光伏元件之視角，探討全光譜吸收材料之設計策略與效能表現間的關聯，聚焦於分子電子結構、電荷轉移特性、界面能階配置，以及電子注入、再生效率與能量損失之間的精密平衡。因此，全光譜吸收光伏材料的分子設計應由單一分子最佳化，轉向分子、半導體與電解質或主動層材料之間的系統性協同，為材料設計由「光譜最大化」邁向「效率最適化」提供具體的概念指引。

## 關鍵詞

染料敏化太陽能電池、有機太陽能電池、分子工程、光收成效率、電荷轉移、鍵長交替、能階匹配、動力學、馬可斯電子轉移理論

## 一、全光譜吸收簡介

太陽光的能譜連續且寬廣，涵蓋紫外、可見至近紅外區（約 300 至 2500 nm）。無論是自然界的光合作用，或人工的光電轉換元件如太陽能電池與光偵測器，其性能皆取決於吸光材料能否有效覆蓋此能量範圍。在自然界中，植物與光合細菌利用由葉綠素、胡蘿蔔素與膽素等高吸收係數（$10^5\,\text{M}^{-1}\,\text{cm}^{-1}$）組成的光捕獲天線（light-harvesting antenna complex），透過精確排列的蛋白質與輔因子達成高效率光吸收與能量傳遞[1]，其能量轉換效率在室溫下可超過 99.98%[2]，顯示自然界對光能利用的高度優化。

然而，要在人工材料中重現自然界般寬廣且高效率的太陽光捕獲並不容易。多數有機或無機錯合物材料的吸收受能隙、π 共軛長度與軌域排列所限制，通常只在特定波長範圍產生強吸收，使太陽光利用受限，進而抑制光電流（photocurrent）的產生與光電轉換效率（photo-to-electricity conversion efficiency，$\eta$）。因此，材料設計逐漸從單波段強吸收拓展至跨越紫外、可見與近紅外光區的「全光譜吸收（panchromatic absorption）」，藉由涵蓋更多電子躍遷以提升太陽能利用率與光伏元件效能表現。

典型的全光譜吸收定義為材料「從紫外區至近紅外光區（400-920 nm）皆呈現高吸收度」[3]，亦即吸光涵蓋紫外、可見與近紅外區域，並具有一定程度的高莫耳消光係數（molar extinction coefficient，$\varepsilon$）。相對於單波段吸收分子，全光譜吸收材料必須同時縮小多組能階之間的能量差，

從而達成不同能量區間的電子躍遷，因此分子設計難度更高。若進一步應用於太陽光伏（photovoltaic）發電，還需同時考量光電流產生機制的各個步驟，以確保有效的光伏效能。本篇文章將從吸收光譜機制出發，探討全光譜吸收材料的分子設計原則與限制。如圖一(a)，光譜中不同波段的吸收主要對應下列電子躍遷：

**(1) 紫外光區（Ultra-violet，UV，<400 nm）：**

主要為定域化電子躍遷激發，包括芳香或共軛雙鍵的 π→π*躍遷與雜原子（如 O、N、S 等）孤對電子對的 n→π*躍遷。前者為電偶極允許躍遷（electric-dipole-allowed transition），吸收強度較大；後者因軌域重疊及躍遷偶極矩（transition dipole moment）較小，屬部分禁止（partially forbidden）躍遷，吸收強度較弱。

**(2) 可見光區（Visible，Vis，400-700 nm）：**

由 π→π*躍遷主導，並伴隨不同程度的分子內電荷轉移（intramolecular charge transfer，ICT 或 CT）。此類激發通常涉及前沿分子軌域（frontier molecular orbitals）間的多重躍遷組合，反映電子激發在共軛主體與側鏈取代基團間同時存在定域化（localize）與離域化（delocalize）的行為，並直接影響主要吸收帶位置與消光係數大小。典型方式如不同比例的 HOMO−1→LUMO 與 HOMO→LUMO+1 激發。其中，HOMO 及 LUMO 分別表示最高已填滿分子軌域（highest-occupied molecular orbitals）及最低未填滿分子軌域（highest-unoccupied molecular orbitals）。

**(3) 近紅外光區（near-infrared，NIR，700-1100 nm）：**

主要源自較低能量的 CT 激發，顯示分子在激發態中具更強的電子離域化與電荷再分佈行為。對於高共軛分子，NIR 吸收通常對應於材料最小能隙的光學躍遷，即 HOMO→LUMO 激發；而在過渡金屬錯合物中，則屬於金屬與配位基間的 MLCT/LMCT（metal-to-ligand/ligand-to-metal charge transfer）激發。

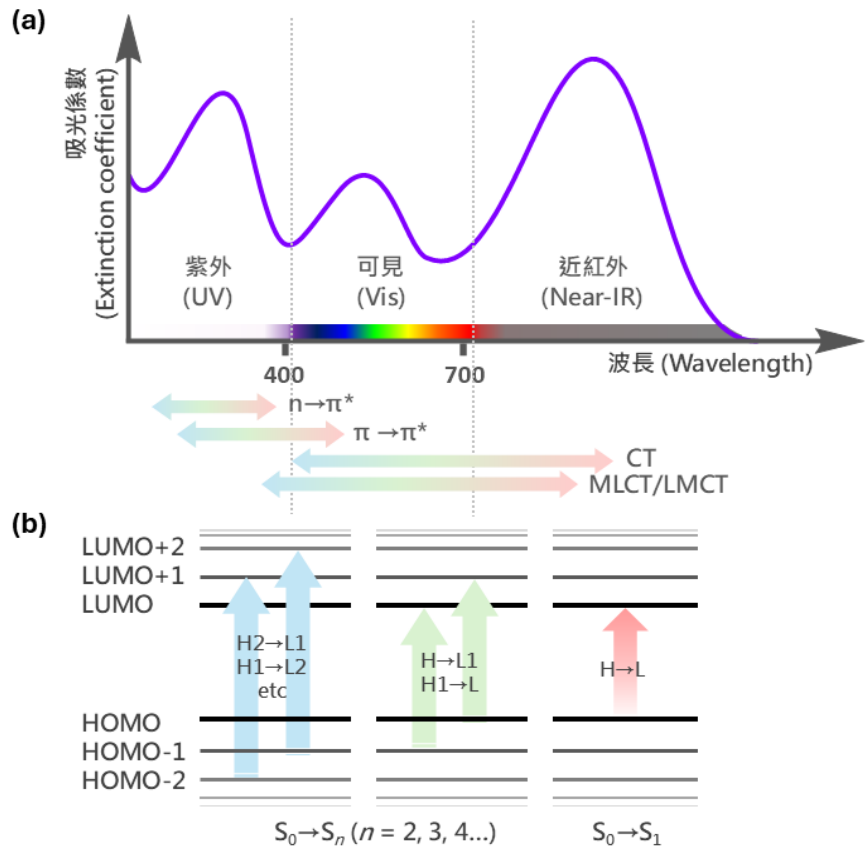

圖一、吸光分子材料的(a)吸收光譜及(b)軌域躍遷激發示意圖

上述不同波段的吸收涉及電子在不同組態間的分子軌域躍遷。如圖一(b)所示，在絕大多數典型的 π 共軛有機分子與金屬錯合物中，最低激發態（$S_0 \rightarrow S_1$）通常完全或主要為 HOMO→LUMO（簡稱 H→L）的躍遷組成。其是否含有其他組成的軌域激發，取決於分子對稱性、共軛程度、推拉電子基效應，以及金屬-配位基間的軌域混成程度而有所不同。此外，最低激發吸收的可見度取決於躍遷偶極矩（transition dipole moment，μ）的大小，因其直接反映吸收強度的高低。當 μ 值過小，吸收峰將十分微弱，往往難以於光譜中辨識，或被相鄰的較強吸收峰掩蓋。次低激發態（$S_0 \rightarrow S_2$）通常包含主要的 H→L1 及 H1→L 躍遷（L1 與 H1 分別指 LUMO+1 及 HOMO−1）。H→L1 及 H1→L 躍遷也同時或分別出現在 $S_0 \rightarrow S_n$（n ≥ 3）激發態，實務上，這同樣視分子結構、共軛程度或推-拉電子效應而定。隨著激發階數提高，$S_0 \rightarrow S_n$（n ≥ 3）之組成更為複雜，可能包含 H2→L1、H1→L2、H→L3、H3→L，以及前述 H→L1 與 H1→L 等多種組合。這些激發往往更具定域化特徵，導致吸收強度相對更高，其對應的消光係數（extinction coefficient，ε）可達數萬至數十萬 $M^{-1}cm^{-1}$。

## 二、全光譜吸收對太陽能電池效能的影響

太陽能電池效率可用以下公式描述[4]：

$$\text{PCE} = \frac{\eta_{out}}{\eta_{in}} = \frac{J_{mp} \times V_{mp}}{\eta_{in}} = \frac{J_{SC} \times V_{OC} \times FF}{\eta_{in}} \qquad (1)$$

其中，$\eta_{in}$為電池接受入射光照之功率，$\eta_{out}$為電池輸出之最大功率，亦即在此入射光照下最大電流密度（$J_{mp}$）與最大電壓（$V_{mp}$）之乘積。而由於電池在開環狀態可有最大電壓，即開環電壓（$V_{OC}$，open-circuit voltage），其與閉環狀態之最大值—短路電流密度（$J_{SC}$，short-circuit current density）和填充因子（FF，fill factor）的乘積，也可用來求得最大輸出功率。此外，在 AM 1.5 G 的標準光照條件下，太陽能電池的短路電流密度可用下式表示[4,5]：

$$J_{SC} = e \int I_0(\lambda) \times IPCE(\lambda) \, d\lambda \quad (2)$$

其中，$e$為基本電荷，$I_0(\lambda)$為 AM1.5 下的入射光子通量，$IPCE(\lambda)$是一種入射光子在外電路中產生電子的外部量子效率，為各入射單波長光子所轉換為電子的轉換效率，即入射光通量與光電流密度的轉換比。實務上，我們可量得電池在不同波長產生對應的電流，並從光譜中可量得電流的最短波長（光子能量最高）處一路積分至最長波長處（光子能量最低），求得$J_{SC}$的上限值。此處，可產生電流的最長波長對應到材料的能隙（energy gap，$E_g$），也與材料的最大吸收峰直接相關，因此能隙越小將有利於更長波長、更寬廣的光吸收範圍，進一步幫助得到越大的短路電流值。以能隙為 1.1 eV、相當於吸光波長約 1,130 nm 的矽晶電池為例，其在 AM 1.5 G 的標準光照下可有 43.6 mA cm$^{-2}$ 的$J_{mp}$輸出[6]。

在太陽能電池中，$IPCE(\lambda)$可拆解為各個相關步驟的效率乘積，不同的電池技術所包含之相關組成也隨之有異。例如在有機太陽能電池（organic solar cells，OSCs）中，$IPCE(\lambda)$可分解為光收成效率（light-harvesting efficiency，$LHE(\lambda)$）、激子擴散效率（$\eta_{diff}$）、由激子至自由載子之解離率（$\eta_{dis}$）、電荷傳輸效率（$\eta_{tr}$）和電荷收集率（$\eta_{col}$）的乘積，如下式(3)所示[4]：

$$IPCE(\lambda) = LHE(\lambda) \times \eta_{diff} \times \eta_{dis} \times \eta_{tr} \times \eta_{col} \quad (3)$$

而在染料敏化太陽能電池（dye-sensitized solar cells，DSSCs）中，$IPCE(\lambda)$則與光收成效率（$LHE(\lambda)$）、電子注入效率（$\eta_{inj}$）、電荷收集率（$\eta_{col}$）、染料再生率（$\eta_{reg}$）等相關，如下式(4)所示[5]：

$$IPCE(\lambda) = LHE(\lambda) \times \eta_{inj} \times \eta_{col} \times \eta_{reg} \quad (4)$$

觀察以上式子可知，$LHE$是太陽能電池進行吸光發電的起手式，也與材料設計息息相關。影響材料$LHE$的因素，主要在於其光譜響應範圍、莫耳消光係數（molar extinction coefficient，ε）和材料於電池中的蘊含量。其中，材料含量的意義隨電池技術不同而異，如 OSCs 決定於主動層的材料添加比例及厚度，或 DSSCs 依賴染料的吸附量等。在未考慮內部損失的情況下，$LHE$可從以下公式近似[7]：

$$LHE(\lambda) = 1 - 10^{-A(\lambda)} \quad (5)$$

這裡，$A$指的是吸光度（absorbance）。在分子溶液狀態下，吸光度$A$可依照比爾定律（Beer-Lambert Law）來表示，如下式(6)；

$$A(\lambda) = \varepsilon(\lambda)bc \quad (6)$$

其中，$\varepsilon$為莫耳消光係數（$\varepsilon$，單位 $M^{-1}cm^{-1}$）；$b$為吸收光徑寬度（單位 dm）；$c$為濃度（單位 M）。而如在固態薄膜條件下，例如染料薄膜吸附二氧化鈦半導體，則吸光度可表示為下式(7)：

$$A(\lambda) = \alpha(\lambda)d \qquad (7)$$

此時，$\alpha$指的是吸收係數（absorption coefficient，單位 $cm^{-1}$）；$d$指的是奈米晶體薄膜的厚度。由於吸光度$A$隨波長而改變，係因分子能階間的能量差距而來，因此不論在溶液或薄膜情況下，分子吸光性質的調控，可高度透過材料分子設計來掌握。此方向的正面實例之一，即為在 DSCs 中使用釕金屬錯合物染料作為光敏化劑（photosensitizer），來敏化 $TiO_2$ 奈米半導體薄膜，只需 6 μm 的厚度，其最大吸收波長附近的$LHE$值就可超過 90%。這也是 $TiO_2$ 奈米薄膜只被吸附單層染料，就可呈現深色的原因。

另一個重要的太陽能電池參數是$V_{OC}$，以 DSSCs 為例，可透過以下方程式來定義[5]：

$$V_{OC} = E_{CB} + kT \ln\left(\frac{n_c}{N_{CB}}\right) - E_{redox} \qquad (8)$$

其中，$E_{CB}$為 $TiO_2$ 導帶下緣（conduction band maximum，CBM）的能階，見圖二；$k$為波茲曼常數，其值約為$1.38 \times 10^{-23}\ m^2\ kg\ s^{-2}\ K^{-1}$；$T$為溫度；$n_c$為 $TiO_2$ 表面的電子密度；$N_{CB}$為 $TiO_2$ 中的有效態密度（density of states）。此外，用以下方程式也可求得$V_{OC}$[8]：

$$V_{OC} = \frac{kT}{e} \ln\left(\frac{J_{inj}}{k_{rec}n_{c,0}[I_3^-]}\right) \qquad (9)$$

其中$J_{inj}$是注入電子的通量，$k_{rec}$是還原$I_3^-$電解質的速率常數，$n_{c,0}$是在不照光狀態下的導帶電子密度。由以上定義可知，$V_{OC}$的理論最大值取決於 $TiO_2$ 的費米能階（Fermi level，$E_{F,n}$）與電解質氧化還原電位（$E_{redox}$）之間的能量差。而前者又取決於$E_{CB}$、$n_c$、$\eta_{inj}$和電荷再結合過程的多重影響。為此，過往已發展許多策略來提升$V_{OC}$，諸如採用具低$E_{redox}$的合適電解質[9,10]、加入添加劑 4-叔丁基吡啶（tert-butylpyridine，TBP）於電解質中以拉高$E_{CB}$[11–13]、引入長鏈烷基至染料分子結構來抑制電荷再結合而增強$n_c$[14–16]等。

由上述說明可知，材料分子的吸光性質與光收成效率（$LHE$）密切相關，並直接影響太陽能電池的短路電流密度（$J_{SC}$）。表面上看來，似乎只要透過分子設計以最大化吸光能力，即可獲得高$J_{SC}$的高效率太陽能電池。研究指出，假設一個 IPCE 為 100%的理想 OSC 元件，其吸光材料對長波長區的光譜響應如達到 700 或 800 nm，理論上$J_{SC}$最大值可分別達到 20.6 和 27.4 $mA\ cm^{-2}$之譜[17–19]。然而，實際上的結構-性質關係並非如此直觀。分子的吸收行為牽涉能階結構、電荷轉移動力學與界面匹配等多重因素，其間具有高度的連動性與權衡關係[20,21]，尤其在全光譜吸收材料的設計上，電子結構的微小改變往往牽一髮而動全身，對整體光電轉換行為造成深遠影響。根據馬可斯電子轉移理論（Marcus electron transfer theory）[22,23]，若分子吸收範圍過寬通常意味能隙極窄，LUMO 能階可能偏低，從而降低介面間電子轉移的自由能驅動力（driving force，$\Delta G$），在馬可斯理論的正常區（normal region，$-\Delta G < \lambda$）條件下，此種驅動力下降將導致電子或電荷傳遞速率（$k_{et}$）減慢，最終使光電轉換過程中界面電子注入（或激子

分離）與再生效率受限，進而表現出較低的 IPCE 及其他光電效能參數。在 OSCs 中，此現象主要反映在$\eta_{dis}$和$\eta_{diff}$的降低，而在 DSSCs 中則影響了$\eta_{inj}$和$\eta_{reg}$，最終抑制整體的光電流表現與量子效率[24]。因此，精密設計具有適當電子結構特性與分子性質之低能隙全光譜吸收材料，對實現令人滿意的$J_{sc}$和 IPCE 至關重要。以下將從分子結構與電子結構等層面，進一步探討其設計原則與影響機制。

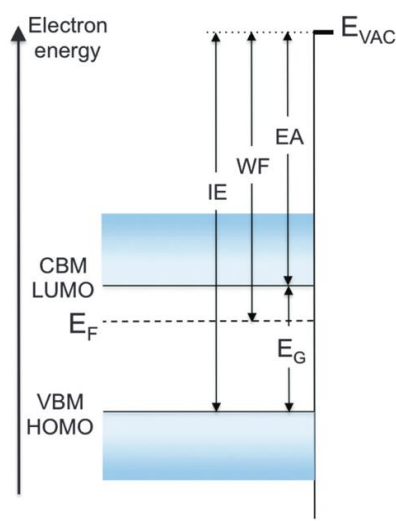

**圖二、**有機及無機半導體材料的能階圖[25]。其中 CBM 及 VBM 分別為價帶及導帶邊緣；HOMO 及 LUMO 分別為最高已填滿及最低未填滿分子軌域；$E_F$或$E_{F,n}$為費米能階；IE 及 EA 分別為游離能及電子親和力；WF 為功函數；$E_g$為能隙。

### 三、全光譜吸收材料的分子工程：

想要提升分子的吸光能力，關鍵的分子設計策略是利用鍵長交替（bond length alternation, BLA）的分子結構特性，如圖三所示[26]。BLA 用來描述共軛分子系統中，單、雙鍵交替所造成的鍵長差異。這些單、雙鍵交替的鍵長，原本可因共振效應而趨向一致，然而實際上仍存在一定差距，此差距大小取決於共振穩定能（resonance stabilization energy, RSE）之影響[27,28]——RSE 越大，BLA 因鍵長差異越小而降低，反之則 BLA 增大。因此，較低的 BLA 可使分子的 π 電子在整個共軛系統主鏈上大幅離域化（delocalization），致使電子躍遷所需的能量降低，從而達到分子吸收光譜的紅位移（往長波長移動）並增強吸收度。此外，降低 BLA 不僅能增強光吸收，也提高分子極化率（polarizability），使其對外電場響應更強[29,30]。這些特性與分子的非線性光學（nonlinear optics，NLOs）行為密切相關。因此低 BLA 結構往往有助於提升二階 NLO 效應，如二次諧波產生（second harmonic generation，SHG），或三階效應，如雙光子吸收（two-photon absorption，TPA）所需的分子特性。

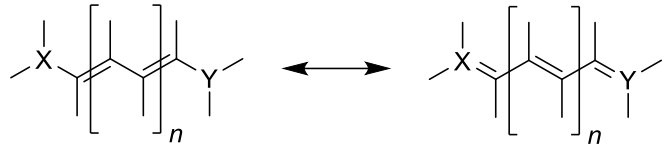

圖三、鍵長交替結構示意圖

  為有效降低 BLA 並增強 π 電子離域化，進而實現吸收的紅位移、光譜拓寬與吸收強度提升，分子設計必須透過延長共軛主鏈、提高芳香環密度、以及導入推-拉電子架構等方式來強化共軛程度，細節將於下一段說明。He 等人以灣位合環化靛藍（bay-annulated indigo，BAI）基團為核心，提出一套可精準調控光譜吸收的二維共軛（2D-conjugation）分子設計策略[31]。此策略結合橫向延展（bay-annulation）與縱向延展（indigo-phenyl extension）兩條共軛軸，分別調控低能量與高能量區的吸收；透過在不同方向引入不同數目的噻吩（thiophene）片段，即可模組化地調整分子吸收行為。如圖四所示，橫向延展系列 **BAI-*n*Tx (*n* = 1–3)**隨噻吩數目增加而能隙逐步縮小，其最低能量吸收峰由 579 nm、660 nm 移至 697 nm，呈現明顯紅位移，對紅外/近紅外光區具有強化效果。另一方面，由 **BAI-*n*Tx** 延伸至 **BAI-*n*Tx-1Ty** 與 **BAI-*n*Tx-2Ty**（*n* = 1–2）的縱向延展則會新增位於 400-500 nm（藍光至綠光區）的高能量吸收峰，使單一分子可同時具備低能量與高能量吸收特徵，形成真正的全光譜吸收。密度泛函理論（density functional theory，DFT）結果清楚支持此結構－性質關係（structure-property relationship）：橫向延展促使 LUMO 更加離域化、有效縮小能隙；縱向延展則提升 HOMO/LUMO 能階並增強躍遷偶極，使高能量吸收顯著增加，且吸收強度隨芳香環延展提升。由於此類分子可同時覆蓋太陽光譜的高、低能量區，因此特別適合應用於 OSCs 與 DSSCs。單一分子即可實現寬光譜吸收，不僅有助於提升$J_{sc}$，也能降低多組成共敏化所帶來的設計與製程複雜度。

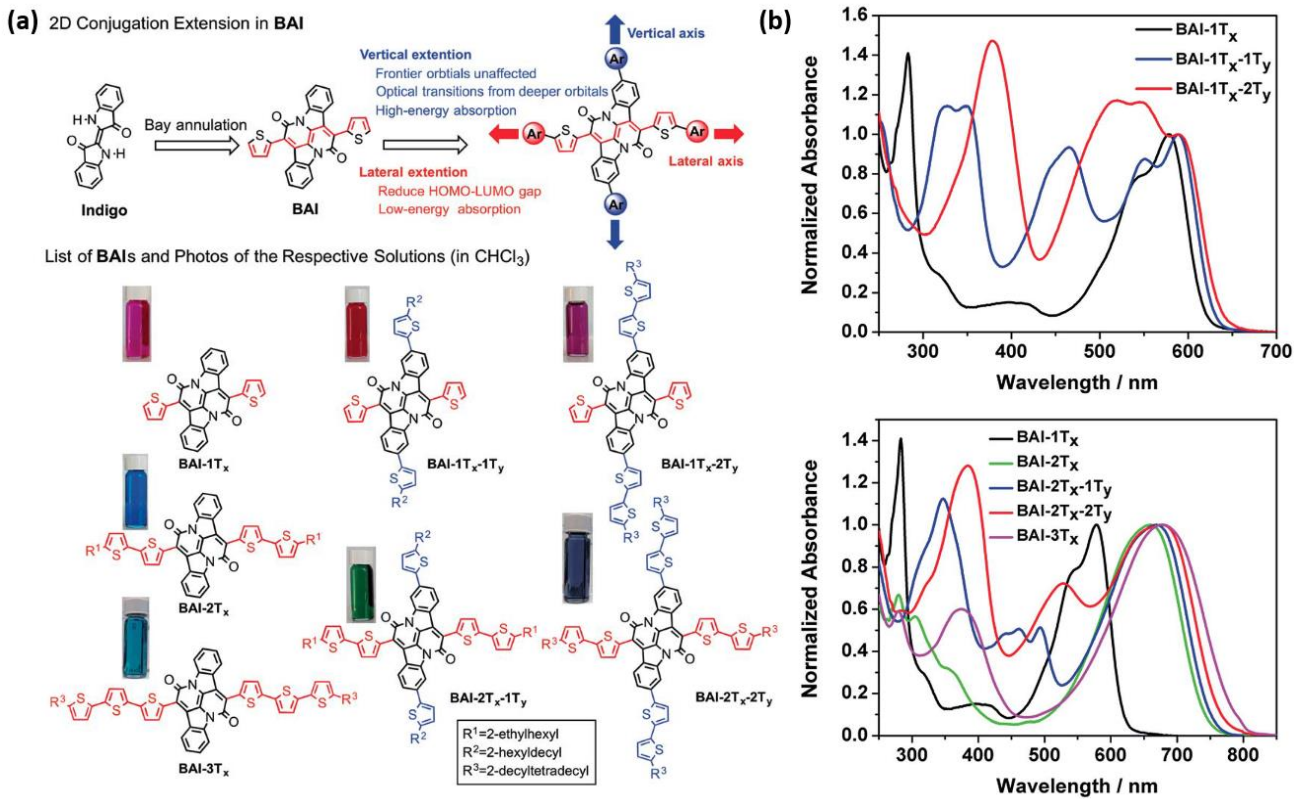

圖四、灣位合環化靛藍衍生物的(a)分子結構與(b)吸收光譜圖[31]。

然而，值得注意的是，單靠 BLA 策略通常只造成近紅外至遠紅外區的光譜吸收，但不足以造成全光譜吸收，亦即吸收光譜無法全面涵蓋紫外、可見及紅外光區。另一方面，過低的 BLA 亦可能引發附帶風險，包括光降解傾向增加、化學穩定性下降，以及開殼態（open-shell）特徵的出現[32]。因此可知，為了達到降低 BLA 的策略，更為謹慎與全盤考量的材料分子設計是必要的，將分別討論如下：

## 1. 推-拉型分子設計

為了提升分子在可見光甚至紅外光區的吸收能力並促進電荷分離，一種常見且有效的策略是採用推－拉型（push-pull type，或稱 D-A type）分子設計[33]。此設計的核心在於於分子主要共軛骨架兩端引入推電子與拉電子基團，分別作為電子予體（donor group，D）及電子受體（acceptor group，A），形成電子密度不對稱分佈的 D–π–A 架構。這樣的設計同時延伸了 π 共軛系統與調控前沿軌域（frontier molecular orbitals）能階，進而使能隙降低。隨著 D–A 間電荷分佈極化增強，分子在激發態中呈現出明顯分子內電荷轉移（intramolecular charge transfer，ICT）特徵，導致吸收光譜紅位移（red shift），甚至延伸至近紅外光區。除了小分子吸光材料外，許多低能隙共軛高分子材料亦採用 D–A 交替共聚結構（alternating copolymer），在主鏈上週期性交替引入 D 與 A 單元，使前沿分子軌域產生高度的能階耦合（orbital coupling）。這種內在的推-拉效應促使電子在共軛骨架上高度離域化，導致能隙縮小與吸收光譜紅位移。此設計策略的主要優勢包括：

(a) 合成策略相對簡單：分子骨架可透過多樣化合成手段實現 π 共軛基團的耦合（coupling）或縮合（condensation）。除了直接官能基化（如溴化、硝化、胺化等）反應外，典型的方法包括連結芳香環彼此或與特定基團之過渡金屬催化耦合反應（transition-metal-catalyzed cross-coupling），如 Suzuki-Miyaura[34]、Stille[35]、Sonogashira[36]、Buchwald-Hartwig[37]等人名反應。此外，亦常利用官能基引導轉換反應，包括活化 α 碳基（–CH$_2$–）與羰基（C=O）的縮合反應（如 Knoevenagel condensation）[38]、形成稠環共軛結構的環化反應（如 cyclocondensation[39] 或 Scholl oxidation[40]）、以及芳香環的氧化耦合（oxidative coupling）反應[40] 或直接芳香基化（direct arylation）[41]等。

(b) 有效達成吸收紅移：設計上常以高度 π 共軛且化學穩定的基團作為分子核心骨架，並於兩端或側鏈進行 D–A 修飾，即可在減少骨架改動的前提下，顯著增強光譜的響應。典型設計平台包括紫質（porphyrin）[5,42]、硼二吡咯甲烷（boron dipyrromethene，BODIPY）[43]、二酮吡咯並吡咯（diketopyrrolopyrrole，DPP）[44]、苯并噻二唑（2,1,3-benzothiadiazole，BTD）[45,46]以及萘嵌苯二醯亞胺（rylene diimide，RDI）[47]等發色團（chromophore），皆可從分子設計有效地調控前沿分子軌域能階，並實現吸收光譜的紅位移與展寬。

(c) 精密能階調控：在 DSSCs 中，染料的最低未填滿分子軌域（lowest-unoccupied molecular orbitals，LUMO）應高於 TiO$_2$ 導帶以確保光生電子（photoelectrons）的順利注入；而最高已填滿分子軌域（highest-occupied molecular orbitals，HOMO）則須低於電解質的氧化還原電位（$E_{redox}$），以利電解質對氧化態染料進行再生[48]；另一方面，OSCs 材料的開發同樣也高度關注 D/A 材料的 HOMO/LUMO 能階[20]。不論早期以富勒烯（fullerene）為受體（A）或現今非富勒烯受體（non-fullerene acceptors，NFAs）為主的高效率系統，激子解離所需的最低驅動力（$\Delta E_{DA} \approx E_{LUMO}^{Donor} - E_{LUMO}^{Acceptor}$）通常位於 0–0.4 eV 之間，既能克服激子結合能（exciton binding energy，$E_b$）而產生電荷解離，又可減少電壓損失（voltage loss，$\Delta V_{loss}$）[49]。因此如前章所述，理想的分子設計除滿足吸收範圍的拓展外，亦需兼顧適當的能階差與介面驅動力，來平衡光吸收與電荷在不同材料與界面間的動力學表現。這類能階調控主要依賴推/拉電子基團的精細設計，微小的取代或結構改變即可重新分配分子內與介面間的電子密度，進而同時影響吸收光譜的紅位移程度與材料之間的能階匹配。

(d) 分子內電荷轉移（ICT）的助益：當 DSSCs 元件中的推-拉型染料吸收光子後，ICT 幫助電子由分子內推電子端激發並遷移至鍵結 TiO$_2$ 表面的拉電子錨基（anchoring group，如 –CO$_2$H、–PO$_3$H$_2$ 等），進而促進激發態電子至 TiO$_2$ 導帶的注入效率、有利於光電流（photocurrent）。同時，分子內另一側的推電子基團也可穩定激發後形成的正電荷，使遠離 TiO$_2$ 表面。如此可幫助降低電荷再結合（charge recombination）的可能性[50]，也有助於電解質對氧化態染料的再生（dye regeneration）。至於在 NFAs 為主的 OSCs 元

件中，適度的 ICT 可產生預極化（pre-polarized）或部分電荷分離（partially charge-separated）的激子（exciton），增強激子偶極距（exciton dipole）並降低 $E_b$，促進 CT 態的形成與電荷分離；但若 CT 過度定域化，則可能提升束縛並抑制自由載子（free charge carrier）生成。[51–53]。因此，適度的 ICT 將有助於界面電荷分離與提升 $J_{SC}$。

然而，推－拉型分子設計也存在潛在挑戰：

(a) 能階帶來的不穩定：為了獲得更大紅位移，往往需要引入強推/強拉電子基團，可能導致 HOMO 能階過高/LUMO 能階過低，進而降低分子的氧化還原與化學穩定性。

(b) 吸收強度的下降：較強的推/拉電子基團雖能增強 ICT，但同時會削弱躍遷偶極矩（μ），造成吸收峰的消光係數（ε）下降。

(c) 能隙理論的限制：當分子能隙因強 ICT 而縮小時，基於能隙理論（energy gap law），激發態與基態間的 Franck-Condon 重疊增加，將造成非輻射鬆弛速率（non-radiative relaxation rate，$k_{nr}$）的提升[54]。此現象可能削弱分子的光穩定性，並影響其成膜性與長期操作穩定度。

關於推－拉型分子設計，文獻中已有大量研究，以下提供部分代表性的實例介紹。Watson 等人的研究指出，推－拉型染料 **NL6** 於強拉電子基團噻吡嗪（thienopyrazine，TPz）兩端引入三芳香胺（TAA）推電子基及氰基丙烯酸（2-cyanoacrylic acid）後能產生強 ICT，進一步在 TPz 側鏈引入兩組額外的 TAA 可形成具有多重 ICT 特徵的染料 **JW1**，如下圖五[55]。後者因跨共軛（cross-conjugated）骨架與多重 Donor 特性使 HOMO/LUMO 分佈更離域化，並具有 6 個主要的 π-π*/ICT 躍遷。這使得 **JW1** 的吸收光譜涵蓋 400-730 nm，除了呈現近乎平坦的 panchromatic 分佈，避免傳統單 Donor 染料 **NL6** 的吸收谷外，於 NIR 區也具有可觀的吸收。顯示 TPz 的預芳香性有助於增強 ICT 與吸收展寬。當應用 **JW1** 染料於 DSSCs 中，可達到 19.7 mA cm$^{-2}$ 的 $J_{SC}$ 與 7.4%的 PCE，明顯優於 **NL6**（$J_{SC}$ 為 13.6 mA cm$^{-2}$，PCE 為 5.4%）。此外，**JW1** 元件在無共敏化的條件下，其 IPCE 在 400-750 nm 皆維持 80%以上的全光譜高效轉換。此元件同時具有高度光穩定性，在連續照光 500 小時後僅約有 10%效能衰退，凸顯此多重 Donor 跨共軛設計兼具光譜展寬與長期耐受性優點。

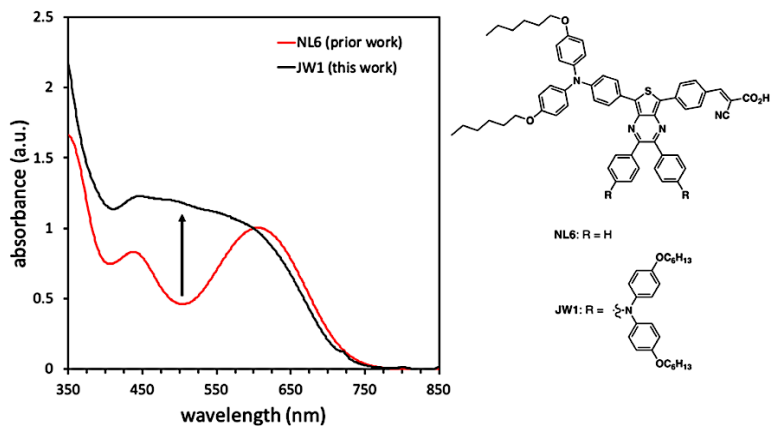

圖五、染料 **NL6** 及 **JW1** 的分子結構及吸收光譜圖[55]。

Ooyama 等人設計了採「雙予體-單受體」的 $D_2$-π-A 結構的推－拉型染料，得到兩種 Y 型結構的染料，如下圖六所示[56]。其分子設計以 BODIPY 為共軛主體，使用二苯胺噻吩咔唑（diphenylamine thienylcarbazole）為予體，並採用吡啶基（2-cyanovinylpyridine）或氰基丙烯酸（2-cyanoacrylic acid）作為受體，而分別得到染料 **OMK-PY** 及 **OMK-CA**。由於分子結構兩側強推－拉電子基團的作用，搭配 π 共軛的延伸使軌域分佈出現明顯 ICT 特徵。因此在吸收光譜上，兩種染料不論溶液中或薄膜吸附於 $TiO_2$ 上，皆展現優異的 UV-Vis-NIR 光捕獲能力，於 300 至 900 nm 區間皆有高莫耳消光係數（最大值分別為 **OMK-PY** 的 95,400 $M^{-1}\,cm^{-1}$ 與 **OMK-CA** 的 83,100 $M^{-1}\,cm^{-1}$）。電化學研究顯示兩種染料皆具有穩定且可逆的氧化還原行為，得益於強大的推拉電子基團。DFT 計算顯示分子 LUMO 主要集中於受體端，有利於電子定向遷移。然而，由於兩種染料的 LUMO 能階（−0.43 至 −0.44 V vs NHE）皆低於 $TiO_2$ 導帶（−0.5 V vs NHE），不利於光生電子的注入，且 $TiO_2$ 表面與電解質界面的電荷再結合也共同限制了光電轉換，因而大幅影響元件效率。最終 **OMK-PY** 及 **OMK-CA** 元件的 $J_{SC}$ 分別為 1.08 與 0.68 $mA\,cm^{-2}$；$V_{OC}$ 分別為 0.34 及 0.32 V；PCE 分別為 0.19% 及 0.11%。整體而言，雙予體 BODIPY 染料雖成功實現高 ICT 與全光譜吸收，但能階匹配與界面電荷再結合也是決定 DSSC 效率的關鍵，這也為後續分子設計提供重要的啟示。

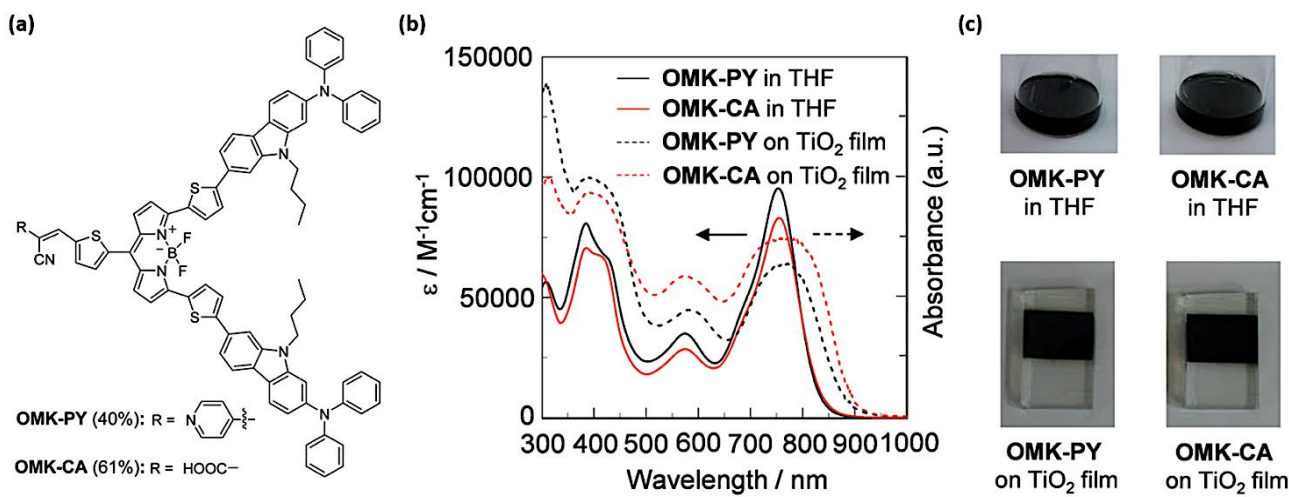

**圖六、染料 OMK-PY 及 OMK-CA 的分子結構及吸收光譜圖及溶液/薄膜態外觀**[56]。

在分子光伏材料中，紫質（porphyrin）因強烈的 Soret band 與 Q-band 吸收而廣泛應用於 DSSC 與 OSC。紫質常被視為人工葉綠素，具備優異的吸光能力，其光譜同時涵蓋 400–500 nm 的強 Soret band（$\varepsilon \approx 10^4$–$10^5$ $M^{-1}$ $cm^{-1}$）與 550–750 nm 的 Q-band（$\varepsilon \approx 10^4$ $M^{-1}$ $cm^{-1}$），是光伏應用中極具代表性的發色團[57]。然而，紫質本身在綠光區（500–600 nm）卻也存在明顯的吸收低谷。苝二醯亞胺（perylene diimide，PDI）與苝單醯亞胺（perylene monoimide，PMI）則具備高強度的綠光至紅光吸收，因此可與紫質形成天然互補。基於此，Lindsey、Bocian 及 Holten 團隊透過橋接炔基將紫質與 PMI 串連而成三元全光譜吸收材料 **T-Ph**，避免了單一染料的光譜缺口[58,59]。然而，透過理論計算（TD-DFT）、暫態吸收光譜（transient absorption，TA）、吸收與螢光光譜等結果，得知其 $S_1$ 激發態屬於定域化激子態（localized exciton state）而沒有形成長距離的電荷轉移態（charge-transfer state），因此無法達到太陽能電池需要的電荷分離（charge separation）步驟。

作者們再利用橋接炔基進一步修飾額外的氫化紫質（或稱氯啉，chlorin）及 PDI，形成十字交叉結構的 **ZnC-T-PDI** 與 **C-T-PDI** 衍生物。有趣的是，由圖七中可知三種多元結構衍生物擁有近乎相同的吸收光譜，然而其電子結構與激發行為卻大相逕庭。由於 LUMO 位於 PDI，HOMO 分布於紫質或 PMI，透過結構中垂直於 T-Ph 共軛方向的 chlorin-porphyrin-PDI 電荷傳輸通道，不含 Zn 金屬配位於 chlorin 的 **C-T-PDI** 只存在 $C-T^+$-$PDI^-$ 的電荷分離過渡態，然而壽命短暫（約 500 ps）。相較之下，含 Zn 配位於 chlorin 的 **ZnC-T-PDI** 可在高極性溶劑中產生量子效率約 30%、壽命達約 1μs 等級的 $ZnC^+$-T-$PDI^-$ 長效電荷分離態，遠優於僅扮演吸光天線（light-harvesting antenna）的 **T-Ph**。因此，**ZnC-T-PDI** 能同時提供強光捕獲與長壽命電荷分離，具備光伏與人造光合作用所需的雙功能反應中心能力。

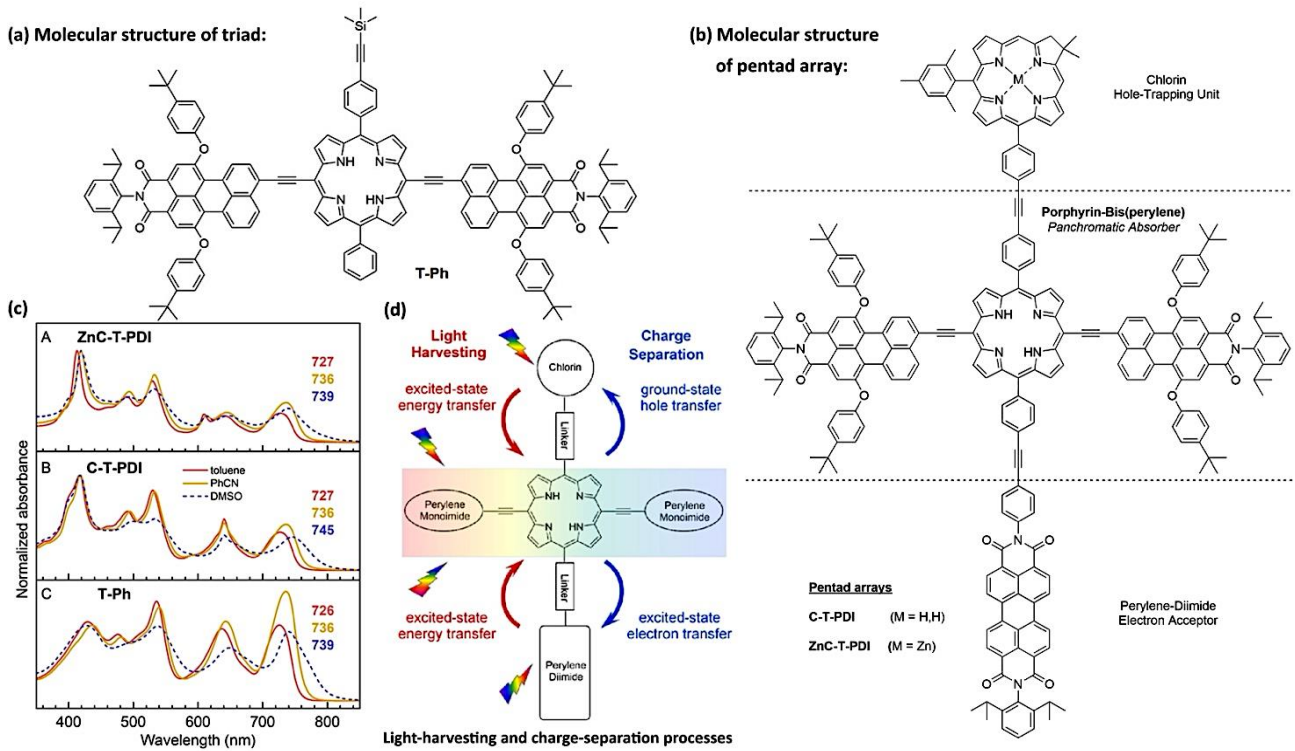

圖七、Chlorin-porphyrin-PDI 及衍生物的(a-b)分子結構、(c)於甲苯、乙腈與 DMSO 中的吸收光譜及(d)光收成與電荷分離機制示意圖[58]。

## 2. 醌式共振結構

另一種分子設計策略著重在共軛骨架中引入具有穩定醌式共振結構（quinoidal resonance structure）的單元[26]，典型的例子是共軛高分子[33,60]與多亞甲基（polymethine）結構[61]。此類骨架多由奇數的亞甲基（-CH=）基團組成，能提供良好的軌域重疊，使鍵長均一化並大幅降低 BLA，形成近似線型的 π 共軛鏈。這種特性不僅有助於提升 π 電子離域程度、也容易達成極窄（<2 eV）的能隙，是實現吸光紅位移與高吸收度的核心機制[26,62,63]。Naim 等人使用花菁（或稱方酸菁，cyanine）類分子 **VG20-C$x$** ($x$ = 2, 8, 16)，視為最典型、穩定也應用最廣的多亞甲基染料之一，可製備高達 75%以上可見光穿夠度（average visible transmittance，AVT）的透明 NIR-DSSCs[64]，如下圖八所示。然而其元件效率僅 3.1%，主因在於紫外至可見光區存在明顯吸收低谷，使將近一半（約 45%）的太陽光能量無法被有效利用；同時，能隙縮小亦導致電子注入與染料再生的過電位（overpotential）不足，使注入過程與由單體向聚集體的非產能能量轉移產生競爭。此一例子清楚顯示，醌式共振結構雖能帶來顯著的紅移吸收，卻同時暴露可見光區吸收不連續、形成吸收缺口，因而未能實現真正的全光譜吸收。同時，窄能隙也帶來染料與材料間電子傳遞受阻的限制。這也凸顯出分子結構設計需在光譜覆蓋完整性、能隙縮小與材料匹配之間取得平衡。

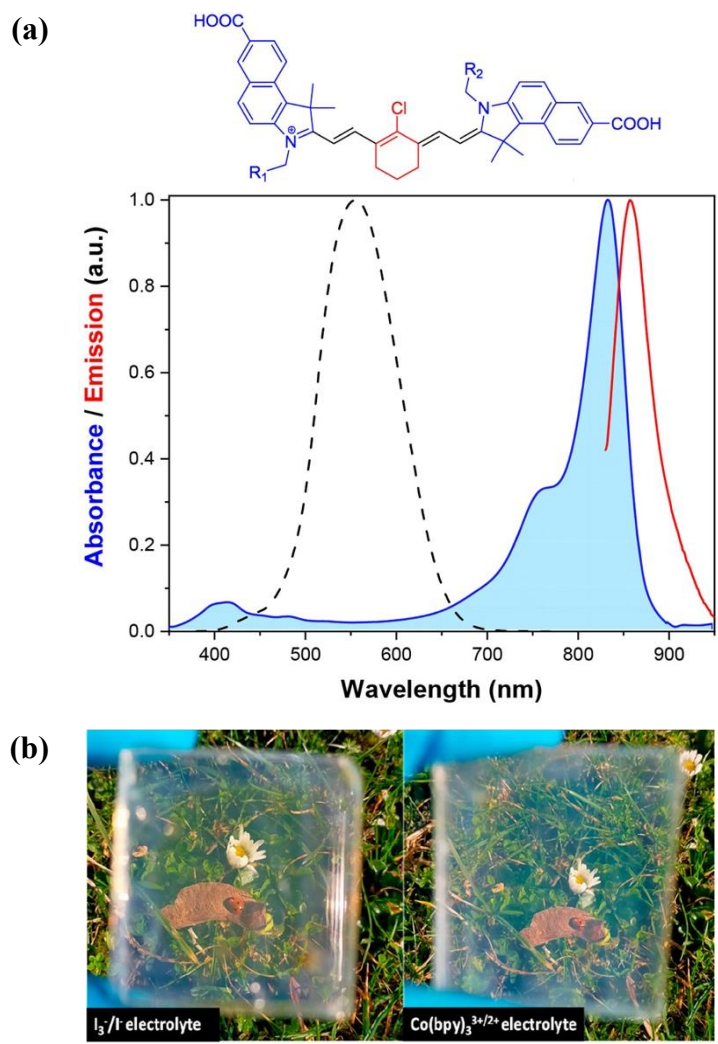

圖八、染料 **VG20-C*x*** 的(a)分子結構與吸收光譜圖；(b)搭配不同電解質之 DSSC 元件外觀[64]。x = 2、8、16 分別代表 $R_1/R_2$ = $C_2H_5$、$C_8H_{17}$、$C_{16}H_{33}$。

欲實現更均勻且寬廣的吸收，其中一種有效的策略為採用 D/A 交替共軛架構，藉由多重電子躍遷與能階分佈來填補光譜缺口。Osaka 及 Takimiya 團隊設計並比較兩種以雙噻吩（thiophene）取代之芳香環并噻二唑為核心的 D-A 共軛聚合物 **PNTz4T** 及 **PBTz4T**，系統性探討其物性與光電應用效能[65]。兩者皆採用近乎共平面的強拉電子性發色團作為受體單元，分別為萘并雙噻二唑（naphthobisthiadiazole，NTz）及苯并噻二唑（benzothiadiazole，Btz）。如圖九所示，DFT 計算顯示 **NTz2T-Me** 單體的 HOMO/LUMO 能階與能隙（−5.52/−3.36 eV，$E_g$ = 2.16 eV）皆略低於 **BTz2T-Me**（−5.40/−2.98 eV，$E_g$ = 2.42 eV）。此結果顯示，除 π 共軛主鏈由苯環延伸為萘環之外，額外引入的噻二唑取代亦有助於穩定醌式共振結構、降低 BLA，從而有效縮減能隙[66]。基於此單體之聚合物 **PNTz4T** 及 **PBTz4T** 皆呈現高剛性、平面性佳與良好

熱穩定性的 D/A 交替主鏈結構，並展現跨越 UV-Vis-NIR 的寬光譜吸收與極低光學能隙。其中，薄膜態的 **PNTz4T**（725 nm，$E_g \approx 1.54$ eV）吸收峰明顯較 **PBTz4T**（662 nm，$E_g \approx 1.65$ eV）為佳，溶液態亦同。當與 $PC_{61}BM$ 摻混製備為 OSC 元件，其最佳表現為 **PNTz4T** 的~6.3%（$V_{OC}$ = 0.76 V；$J_{SC}$ = 12.0 mA cm$^{-2}$；FF = 0.69），顯著優於 **PBtz4T** 的~2.6%（$V_{OC}$ = 0.74 V；$J_{SC}$ = 5.6 mA cm$^{-2}$；FF = 0.63）。此效能差異主要歸因於 **PNTz4T** 更強的吸光能力、更緊密的 π-π 堆疊，與更高效的載子收集行為，顯示 NTz 單元是高效 D-A 共軛聚合物設計中的關鍵骨架。

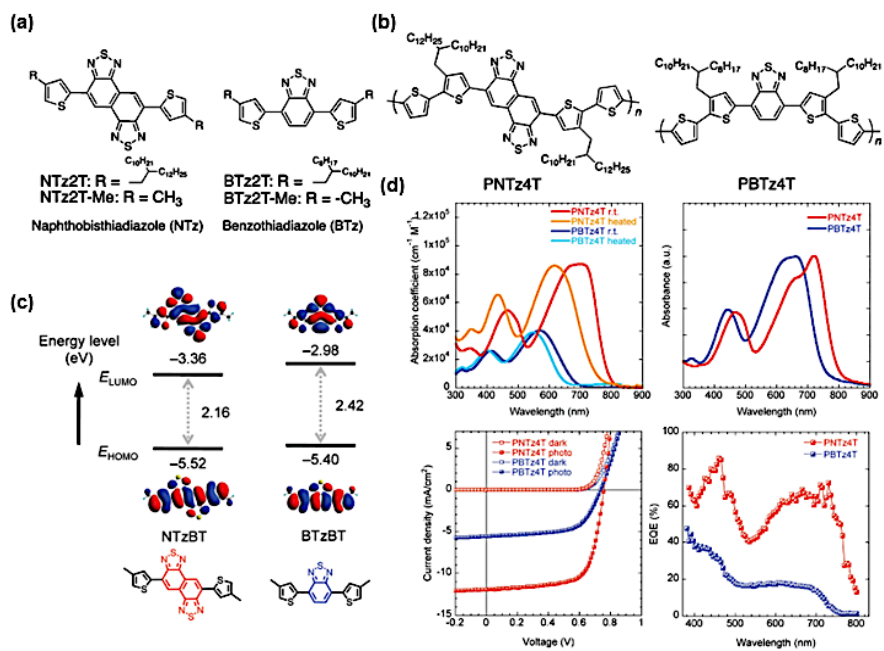

圖九、含芳香環并噻二唑之(a)單體及(b)聚合物的分子結構；(c)單體之前沿分子軌域圖；(d)聚合物之溶液與薄膜吸收光譜圖、J-V 曲線圖及 IPCE 光譜圖[65]。

對於高效率 OSC 而言，予體 HOMO-受體 LUMO 的能階差是否夠小，與材料帶隙是否接近 Shockley–Queisser 理論最佳值（≈ 1.34 eV）皆至關重要。前者可透過合理的 HOMO/LUMO 匹配來有效降低電壓損失（$\Delta V_{loss}$），後者則代表材料能更充分捕獲太陽光。例如，**PDTTDPP** 是由醌型結構單元 二酮吡咯并吡咯（diketopyrrolopyrrole，DPP）和芳香基團二噻吩并噻吩(dithienothiophene，DTT)所組成，是具有 NIR-II 活性的典型高分子[67]。在分子共軛主鏈因 DPP 及 DTT 單元的耦合而產生高度電子離域化。因此，**PDTTDPP** 的能隙僅約 1.22 eV，常用作 OSC 和有機光偵測器（organic photodetector，OPD）元件的予體材料。另一方面，更高的載子遷移率（carrier mobility）亦對抑制再結合、提高電荷收集至關重要。以上三項因素共同構成高效率 OSC 的核心條件。然而，以往高效率材料多吸收到 800 nm 以內，且厚膜>100–150 nm 時容易因傳輸距離變長，導致再結合的大幅增加而影響 FF 與 $J_{SC}$，對厚膜與大面積製程不利。

研究人員近年發現，窄帶隙、強 NIR 吸收且能階可精密調控的非富勒烯受體（non-fullerene acceptors，NFAs）是同時提升 $J_{sc}$ 與 $V_{OC}$ 的關鍵。2019 年，**Y6** 受體的問世與搭配 **PM6** 予體使單接面二元混摻（single-junction binary-blend）的 OSC 元件效率突破至 15.7%，具備劃時代的意義[68]，如圖十所示。近代的高效率 OSC 研究廣泛使用 **PTB7-Th**、**PBDB-T** 與 **PM6** 等 P 型導電高分子為低能隙材料（$E_g \approx 1.3\text{-}1.7 \text{ eV}$），搭配具備強吸收、可調電子結構、高激子離域化與低 $\Delta V_{loss}$ 等優點的非富勒烯受體（如 **ITIC**、**Y6** 等）[69–71]，促使後者迅速取代富勒烯成為當前最具研究與實際應用潛力的材料組合。廣義的低能隙 **Y6** 家族受體與高性能 **PBDB-T** 高分子家族（即 **PM6** 等）予體的組合[72]，搭配結晶調控添加劑（crystalline modifier），可進一步優化三元（tertiary-blend）主動層形貌，使如 **D18-Cl:N3:AT-β2O** 之元件 PCE 突破歷史新高，來到 20.82%[73]。這些材料普遍具備醌式共振結構特徵：針對 P 型導電高分子予體，其主鏈多採 D/A 交替共軛結構，可有效降低 BLA、增強 π 電子離域化，使吸收帶向紅光甚至近紅外光延伸；而針對 NFA 受體則常在強拉電子的共軛主鏈上引入額外的受體基團，形成 A-D-A'-D-A（或 A-DA'D-A）型小分子架構，以近一步穩定醌式共振。整體而言，穩定化的醌式共振是提升光譜利用率與高光電轉換效率的關鍵分子設計策略。

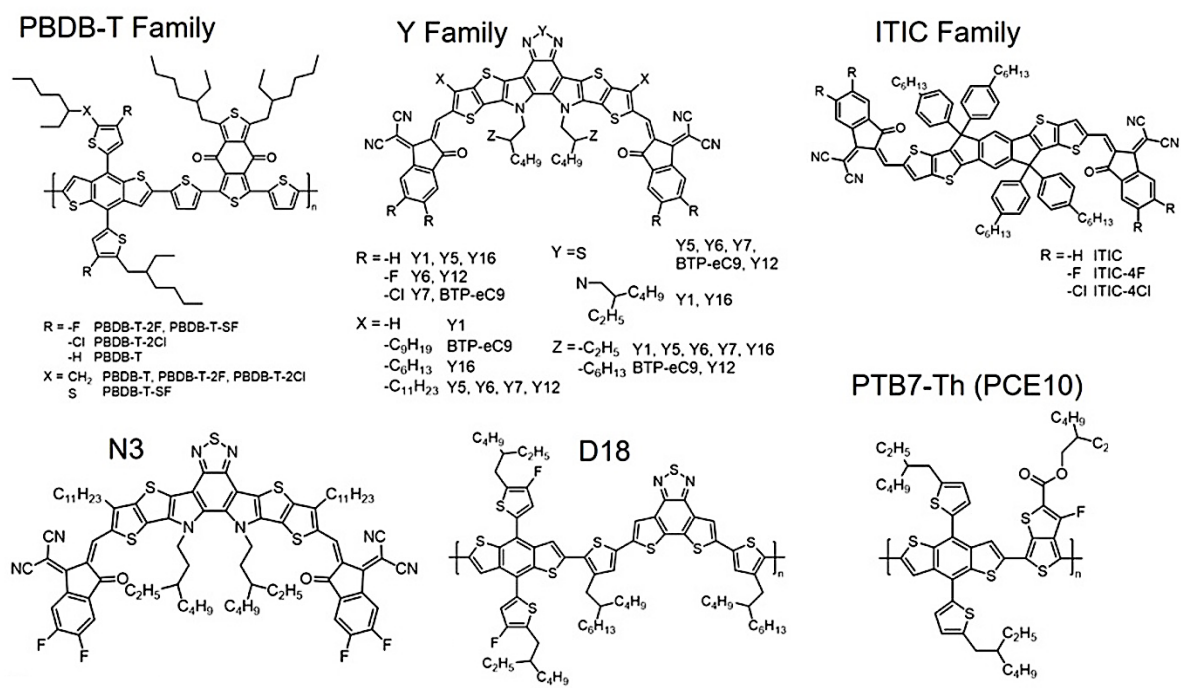

圖十、近年典型 P 型及 N 型 OSC 材料的分子結構[49]。

Yuan 等人所開發的新型 NFA 材料 **Y6** 採用 A-DA'D-A 的分子架構（見圖十），其結構中心為 DA'D 融合環單元（dithienothiophen[3,2-*b*]-pyrrolobenzothiadiazole），兩端則修飾強拉電子性 2FIC（2-(5,6-difluoro-3-oxo-2,3-dihydro-1H-inden-1-ylidene)malononitrile）基團，藉由結構中心

輕微扭轉與兩端平面化的設計，使分子在維持良好共軛的同時避免過度聚集[68]。因此，**Y6** 於薄膜狀態呈強烈的 NIR 吸收（$E_g^{opt}$ = 1.33 eV，吸光啟始值 ≈ 931 nm，ε = 1.07 × 10⁵ cm⁻¹），但仍保有藍綠光區基本的吸收強度，且 HOMO/LUMO 能階約為−5.65/−4.10 eV。這些數據皆與具可見光吸收的 **PM6** 形成良好吸光互補及能階匹配，有助降低能量損失並維持較高的$V_{OC}$。此外，**Y6** 也具有明顯平面朝上（face-on）的 π–π 堆疊，其間距 d ≈ 3.57 Å，有利於與 N 型材料 **PM6** 摻混後仍保有連續的電子傳輸通道，進而提升激子解離與電荷收集。最終，最佳化 **PM6:Y6** 元件於 150 nm 厚度下可達 PCE 15.7%（$J_{SC}$ = 25.3 mA cm⁻²；$V_{OC}$ = 0.83 V；FF = 74.8%），其 EQE 在 450–830 nm 皆維持 70-80%的高效率，即使厚度達 300 nm 仍保持>13%效率，展現優異的厚膜容忍度與光生載子的利用率。此結果充分展現 **Y6** 因醌式共振結構帶來的窄能隙、深 LUMO 能階及良好的 π 堆積，奠定後續 Y 系列受體突破 20%以上 PCE 的材料基礎。

## 3. 官能基化多環芳香烴

典型的多環芳香烴（polycyclic aromatic hydrocarbons, PAHs）是由兩個或多個芳香環以稠環（fused-ring）方式組成，也可能加入具有雜原子的五員或六員環[60,74]。為了達到降低 BLA 的目的，PAHs 不同於醌式共振結構採加強特定芳香環間彼此鍵結的作法，而是納入整個芳香環為均勻 π 共軛系統，從而減少定域化 C=C 鍵的存在，實現更低的 BLA 與更高程度的電子離域化。因此，雖與聚次甲基型（polymethine）分子同樣建構在高度 π 共軛特性上，但 PAHs 結構的高度平面化使電子能大幅離域化在整個分子骨架，加上極佳的疏水性與熱穩定性，使其成為設計全光譜吸收材料的重要類型。隨著共軛芳香環數量增加，PAHs 材料的電子在分子骨架中的離域程度提升，使 HOMO 能階上升、LUMO 能階下降，導致能隙縮小，典型的例子如萘嵌苯二醯亞胺（RDI）[75,76]。當進一步引入推/拉電子基團，或於骨架中摻入 N、O、S 等雜原子，除可延伸 π 共軛系統、調控分子內電荷分佈與能階位置外，還能促使能隙更加收縮（約 2.5 eV 降至 1.5 eV 以下），增強 ICT 特徵而達成顯著紅位移與高消光係數的吸收，同時維持優異的光動力學與載子傳輸性能[77]。

雖然提高平面性與共軛延伸能有效縮小能隙並推進吸收，但也容易導致強烈 π–π 堆疊、分子聚集與能階過度下降。如下圖十一所示，Mai 等人系統化比較不同連接方式的推-拉型雙紫質（diporphyrin）染料 **YDD*n***（*n* = 0, 1–3），探討共軛延伸與稠環化對 UV-Vis-NIR 吸收特性及光伏表現的影響[78]。以本研究的單紫質 **YD0** 為例，透過在 *meso* 位以 π 共軛連接兩個紫質單元所得之雙紫質分子，可有效填補綠光區的吸收缺口，並將吸收延伸至近紅外光一區（near-infrared region I 或 NIR-I，700–900 nm），形成更完整的全光譜覆蓋，如 **YDD0** 的 300–800 nm 與 **YDD1** 的 300–700 nm。相較之下，採用 *meso*-β 位稠環方式建構的融合型雙紫質則具有更強的 π 共軛與電子偶合，使吸收推進至近紅外光二區（near-infrared region II 或 NIR-II，900–1,700 nm），且呈現明顯分裂與展寬的 Soret band，如 **YDD2**（300–1,250 nm）與 **YDD3**（300–950 nm）。

然而，這類融合型結構雖延伸吸光至 NIR-II，卻也犧牲了 NIR-I 區的高強度吸收，且過度共軛導致能隙降低至約 1 eV，也伴隨 LUMO 能階下降至低於 TiO$_2$ 導帶，造成光電子難以有效注入。上述結構-能階差異直接反映於元件表現。如下圖十一所示，**YDD1** 兼具適度電子偶合與良好能階匹配，使其 DSSC 元件的 IPCE 在可見光區保持高於 70%，$J_{SC}$（約 13 mA cm$^{-2}$）與 PCE（5.2%）皆略優於單紫質 **YD0**（$J_{SC}$ ≈ 13 mA cm$^{-2}$；η = 5.1%）。反之，吸光範圍更寬的 **YDD0** 因結構高度平面化易產生聚集而降低電子注入效率；融合型雙紫質則因 LUMO 過低無法注入電子至 TiO$_2$，其元件 $J_{SC}$ 與 IPCE 值趨近於零。整體而言，此研究明確指出寬光譜吸收並非提升光伏效率的唯一關鍵，唯有同時兼顧能階匹配、有效電子注入與材料物理特性等，才能將吸收優勢轉為實際元件效能。

由上例可知，高度平面的分子結構亦帶來嚴重缺點，包含分子間顯著的堆疊（即 π–π stacking）形成聚集體（aggregates），使溶解度驟降並改變光學性質[79–81]。實務上，在結構中引入扭曲基元（twisted units）、螺環（spiro linkages）、氮雜環（aza-fusion）或內醯亞胺（imide）等結構單元，可在維持共軛性的同時抑制過度 π–π 堆疊而形成聚集，提升分散性與薄膜成膜性。

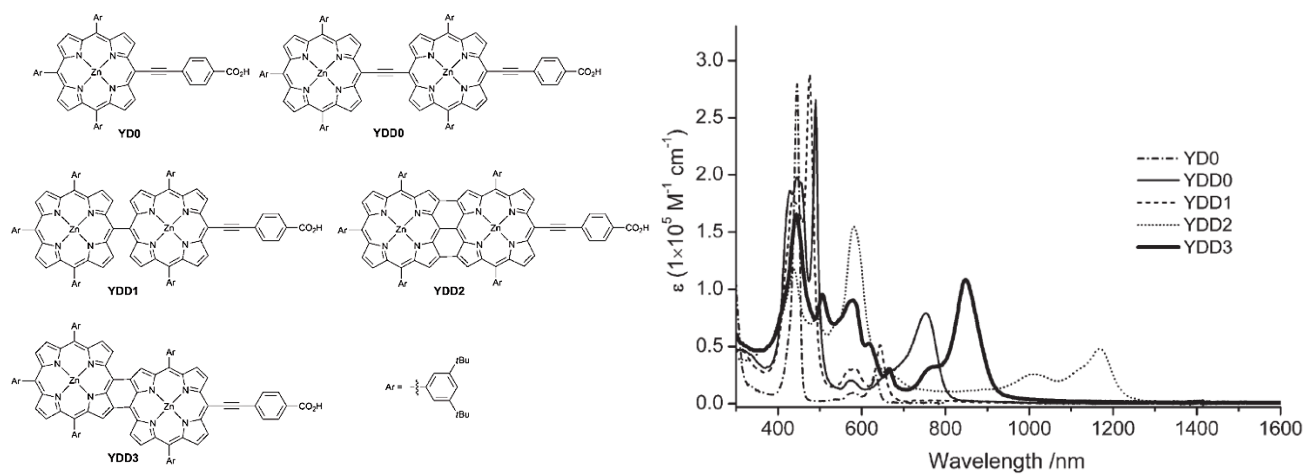

**圖十一**、單紫質與雙紫質染料的分子結構與吸收光譜圖[78]。

萘嵌苯二醯亞胺（RDI）衍生物因具備高度穩定的光物理及電化學性質[83]，被廣泛應用在能源儲存[84,85]、生物技術[86]與光電應用[60,75]等領域。與醌式共振特徵的分子類似，RDI 可透過適當修飾使吸收光譜延伸至近紅外光區；然而，能展現真正全光譜（UV-Vis-NIR）吸收的案例仍相對有限。在 RDI 材料家族中，核心取代之 cRDI（core-substituted，RDI）經由修飾異原子取代基[87]、進行環化（annulated）或稠環化（fused）等在芳香環上的 π 共軛延伸，可形成具高度官能基化的多環芳香烴衍生物，此類分子通常展現優異的寬廣甚至全光譜吸收能力[88]。此外，在 imide 位置與鄰-芳香二胺（o-aryldiamine）衍生物進行縮合-環化反應，可進一步對共軛骨架進行稠環化，形成雙取代之 perinone、PTCBI（3,4,9,10-perylenetetracarboxyl-bisbenzimidazole），或單取代衍生物。此類稠環化 RDI 結構也因高度 π 離域化而展現跨越 UV-Vis-NIR 的寬廣吸收

[89,90]。如下圖十二(a)，如將兩個 PDI 分子進行核心融合而得 t-diPDI 類分子 **18a**，可大幅提升光捕獲範圍，從 425–500 nm 擴張到 300–725 nm[91]。如下圖十二(b)所示，Suraru 等人系統性比較了 **syn-CbDI**、***anti*-CbDI** 與 ***anti*-BfDI**（分別為圖中桃紅色、綠色與橘色曲線）三類核心取代 cNDI 材料的光物理性質，以探討拓撲（syn/anti 連接方式）、取代基電子效應與分子堆疊等對光譜、能階與載子傳輸的影響[92]。其中，具有苯并呋喃（benzofuran）與吲哚（indole）雙稠環取代的 ***anti*-BfDI** 材料具有最大的能隙與約 350−525 nm 的最窄吸光範圍；以同側雙吲哚稠環取代的 ***syn*-CbDI**，其吸收峰大幅紅位移（約 350−625 nm）並呈現明顯 CT 特徵；如兩吲哚以相反方向進行雙取代而得 ***anti*-CbDI**，則可展現跨 400−660 nm 的寬光譜吸收與約 2.0 eV 的最窄能隙。這些結構皆屬於典型 NDI 受體材料，且可藉由核心結構的 N/O 原子置換及吲哚基團上的鹵素/$CF_3$ 修飾，來調控 LUMO 範圍在約−3.1 至−4.0 eV。此外，單晶結構顯示高度平面化的 ***anti*-CbDI** 呈現一維形式的 π 平面堆疊（π-stack），而結構部分扭曲的 ***syn*-CbDI** 則呈現二維方向的分子堆疊（slip-stack）。載子遷移率（mobility）量測顯示 CbDI 材料多雙極性（ambipolar）傳輸特徵，而電子親和力更強的 **BfDI** 與 **$CF_3$-CbDI** 則傾向 N 型傳輸。由此例可具體看出，π 擴張型 NDI 系列材料可藉由拓撲選擇、取代與堆疊方式的精密設計，展現寬光譜覆蓋、能階調控、與 P/N 通道切換的高度材料設計潛力。

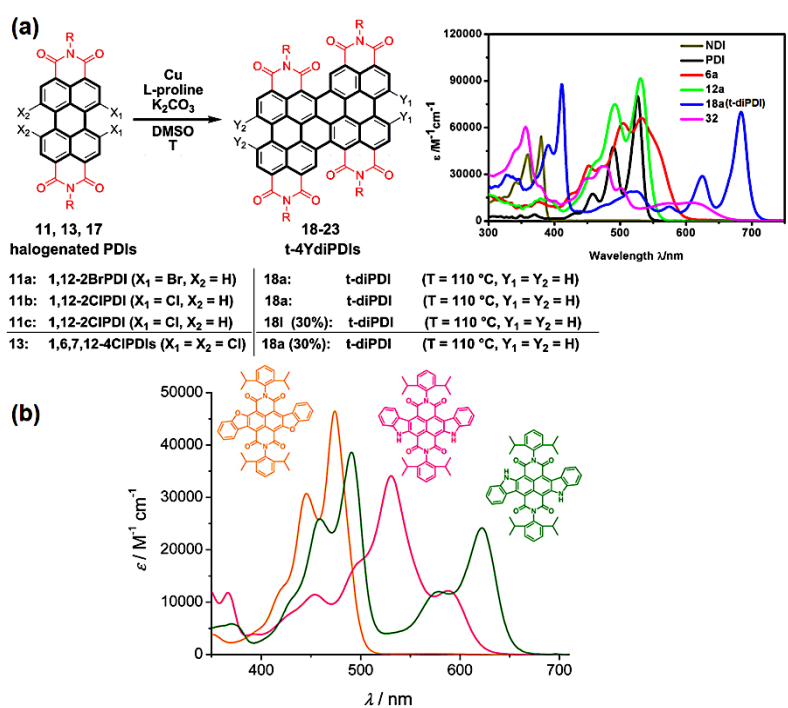

圖十二、(a)融合型 PDI 分子的合成及吸收光譜[91]。途中紅色、綠色、藍色與粉紅色曲線分別代表 **s-diPDI**、**d-diPDI**、**t-diPDI** 與 **NDI-PDI**；(b)核心取代 cNDI 分子結構與吸收光譜[92]。圖中桃紅色、綠色與橘色曲線分別為 ***syn*-CbDI**、***anti*-CbDI** 與 ***anti*-BfDI**。

## 4. 金屬錯合物的電荷轉移吸收

由圖一可知，全光譜吸收的核心在於分子內的電荷轉移（CT）激發，其激發態能量由予體（$E_{donor}$）與受體（$E_{acceptor}$）能階差所主導，即

$$\Delta E = E_{acceptor} - E_{donor} \qquad (10)$$

對於金屬錯合物，CT 躍遷則多為金屬到配位基（metal-to-ligand charge transfer，MLCT）或配位基到金屬（ligand-to-metal charge transfer，LMCT）的電子轉移。以 MLCT 而言，其金屬 d 軌域能量高於配位基 π*軌域，ΔE 縮小得以使 MLCT 躍遷能量降低，吸收可紅位移至可見甚至近紅外光區（約 600–900 nm）。此類吸收多呈現寬廣特徵，對光伏應用材料相當有利，其原因包括軌域混成（orbital mixing）造成多重激發態疊加，以及重金屬中心（如 Ru、Ir、Os 等[93]）帶來的強自旋-軌域耦合（spin-orbital coupling，SOC）而顯著穩定低能量 CT 激發態[94]。這方面的典型例子包括 DSSC 發展早期廣泛使用的釕金屬錯合物染料，如 N3、N719 及 N749（又稱 black dye）[7]，其寬廣的 MLCT 吸收對提升光捕獲能力與光電轉換效率具有關鍵作用[95]；另一方面，過渡金屬酞菁類（metal phthalocyanine，MPc，M 常為 $Zn^{2+}$、$Cu^{2+}$）因能提供強烈的可見-近紅外光吸收，常沉積在 OSCs 和有機薄膜電晶體（organic thin-film transistors，OTFTs）中作為電荷傳輸層（charge transporting layer）[96]。

然而，MLCT/LMCT 的吸光強度（ε 約 $10^2$-$10^3$ $M^{-1}$ $cm^{-1}$）通常不及有機推-拉型分子（ε 約 $10^4$-$10^5$ $M^{-1}$ $cm^{-1}$）。原因在於軌域重疊積分小造成較小的躍遷偶極距（transition dipole moment），加上躍遷通常為部分允許（partially allowed）態而需振動耦合（vibronic coupling）或 SOC 來克服選擇定則（selection rule）的限制。此外需注意的是，金屬中心的氧化還原電位可透過配位基上結構與電子性質的微小調控而顯著改變，可知配位基設計對其氧化還原行為具有強烈影響。這種高度可調性雖有助於分子工程（molecular engineering）應用於全光譜吸收錯合物設計，卻也可能導致錯合物在操作條件下的不穩定性。

Juwita 等人以實驗與 SOC-TDDFT 計算研究重金屬中心的 SOC 如何影響金屬錯鹽染料 [M(tctpy)(X2bpy)Cl][$PF_6$]（M = Os, Ru；tctpy = 4,4',4''-tricarboxy-2,2':6',2''-terpyridine；X = dmbpy or TH-EDOT-bpy）之全光譜吸收與 DSSC 光伏表現[97]。研究顯示，以鋨(II)金屬（Osmium，$Os^{II}$）為中心、TH-EDOT-bpy 為共軛輔助配位基的 **CYC-33O**，因 $Os^{II}$ 的重原子效應使 $^1$MLCT（546 nm，ε ≈ $1.96 \times 10^4$ $M^{-1}cm^{-1}$）與 $^3$MLCT（810 nm，ε ≈ $4.5 \times 10^3$ $M^{-1}cm^{-1}$）躍遷透過強度借用機制（intensity borrowing mechanism）被同步放大。其中，$^3$MLCT 吸收強度較未含長碳鏈 EDOT（3,4-ethylenedioxythiophene）取代的 **Os-3** 增強逾 45%且明顯紅位移，吸收邊緣可達約 950 nm，展現紫外-可見-近紅外光之全光譜吸收。以 **CYC-33O** 製成 DSSC 元件，其 IPCE 值更延伸至 >1000 nm，且在 560 nm 及 870 nm 仍有約 73%與 42%的響應。因此，全光譜之光收成能力使元件$J_{SC}$提升至 19.4 mA $cm^{-2}$，有效彌補鋨錯鹽染料常因較高電子注入驅動力與較快電子傳輸（$\tau_{tr}$短），而伴隨較快電荷再結合（$\tau_{rec}$短），導致$V_{OC}$略低於釕系統的限制，使其最終 PCE 值

（5.7%）仍能與 N749 與 **CYC-33R** 相當。整體而言，本研究展現重原子 SOC 效應與共軛輔助配位基推動全光譜吸收鋨錯鹽染料的高度協同設計。

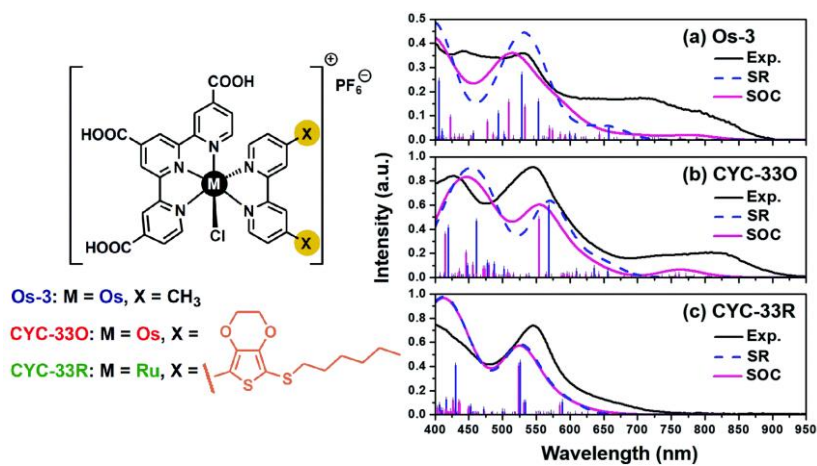

**圖十三**、鋨金屬及釕金屬錯合物的分子結構與實驗、UV-Vis-NIR 吸收光譜圖[97]。

## 5. 共敏化策略與協同夥伴染料

由於設計寬光譜吸收範圍的單一分子具有相當高的設計難度，許多研究者嘗試以「分子雞尾酒（molecular cocktails）」方式，利用離子作用力、超分子化學或材料化學策略，將多支染料混合以拓寬吸收範圍，於 DSSC 應用中此即所謂的共敏化（co-sensitization）策略[98]。藉由同時使用兩種或以上的吸光染料來擴展光譜吸收範圍，可實現模仿太陽光譜的全光譜吸收，從而提升光電轉換效率。

在這方面，最具代表性的成功案例之一即為推-拉型紫質主染料 **YD2-*o*-C8** 與共敏化有機染料 **Y123** 的經典搭配，且曾創下 DSSC 光電轉換效率的世界紀錄[99]，如圖十四所示。**YD2-*o*-C8** 具有強可見至近紅外光吸收，其 *meso* 位苯基引入鄰位長鏈烷氧基，不僅能抑制分子堆疊、提升 $TiO_2$ 表面覆蓋率，也能有效降低電子再結合。搭配氧化還原電位較低、與染料能階高度匹配的 $Co^{2+}/Co^{3+}$ 電解質後，其 **YD2-*o*-C8** 單一染料元件於 450–680 nm 皆能達到超過 80% 的 IPCE。然而，由於紫質本身存在 530 nm 附近的綠光吸收缺口。作者進一步引入吸收峰位於 532 nm 的 **Y123** 作為輔助染料，以主染料:輔助染料為 4:1 的最佳比例共敏化，使元件 IPCE 在 400–700 nm 可達 80–90% 的寬光譜響應。最終共敏化元件可獲得 12.3% 的 PCE、17.7 mA cm$^{-2}$ 的 $J_{SC}$ 與 0.93 V 的 $V_{OC}$。儘管過量 **Y123** 會使 $V_{OC}$ 輕微下降並縮短電子壽命，但互補吸收所帶來的電流增益遠大於這些損失，因此整體效率仍然提升。

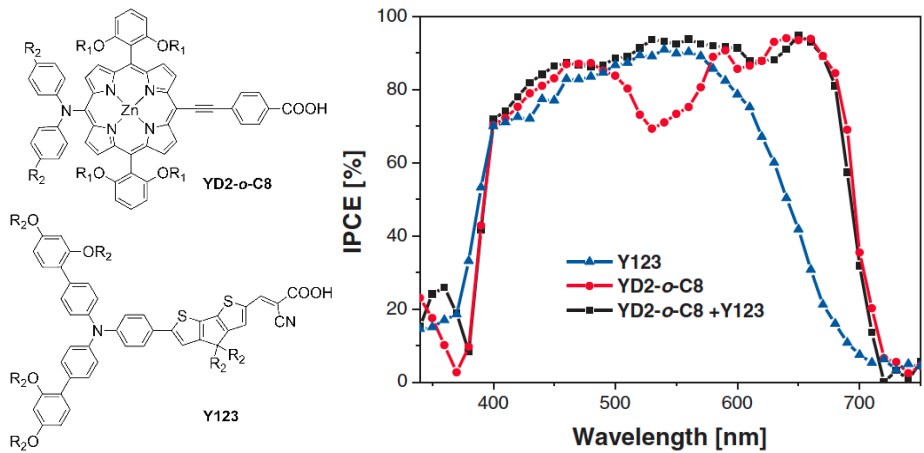

**圖十四、紫質染料 YD2-*o*-C8 與共敏化染料 Y123 的分子結構及 IPCE 圖**[49]。

另一方面，Cooper 等人開發了一套由資料探勘、高通量計算到元件實驗的流程，用於設計全光譜吸收之共敏化 DSSC[100]。透過 ChemDataExtractor 自動擷取 9431 筆染料資料，並以 PM7 半經驗法計算偶極矩、錨基（anchoring group）、光譜覆蓋度等條件來逐層篩選，再以 DFT 驗證能階匹配，最終選出 5 個純有機染料。結果顯示演算法預測的 **XS6＋15** 共敏化系統在 400–700 nm 具有連續吸收與良好光譜互補，其 DSSC 的 IPCE 高達 80-90%，且 $J_{sc}$ 與 $V_{oc}$ 皆較單敏化提升。最終共敏元件效率為 N719 的 92%，相較單敏化染料 **15** 的元件提升約 38%。近一步實驗證實這樣的組合在 TiO$_2$ 表面形成約 19 Å 厚的單層吸附結構，覆蓋率超過 70%，聚集面積明顯減少，對應電子壽命延長並降低了界面電荷再結合。

如上所述，共敏化方法可有效拓寬吸收範圍與提升 $J_{sc}$，甚至在某些情況下改善 $V_{oc}$。然而，這種方法需要克服染料分子間的競爭吸附問題，因此仰賴繁瑣的染料浸漬（dye impregnation）條件優化。在使用酞菁類（phthalocyanine）的共敏化系統中，分子間競爭吸附、聚集與能階錯未適配常導致轉換效率不如預期，甚至低於單一染料元件。相較於分子間共敏化（intermolecular cosensitization）的先天限制，一種新型態的分子內工程方法相應而生，亦即協同夥伴染料（concerted companion dyes）[5]，在分子結構上預先規劃主染料與夥伴染料的協同與互補。具體而言，協同夥伴設計整合兩種吸收特性互補的吸光單元於同一分子中，由於兩單元彼此以飽合碳鏈連接而不共軛，因此吸收行為彼此獨立。此設計不僅可實現單一分子的分子內光譜互補（intramolecular spectral complementation），也同時克服競爭吸附與聚集問題，因此能有效提高效率。

此方面的經典例子如圖十五所示，Zeng 等人開發一系列結合紫質主單元（**XW51**）與 D-A 型有機夥伴單元（**Z2**）的協同夥伴染料 **XW60–XW63**，這些染料分別以不同長度的烷氧鏈串接而成[101]。因紫質特有的 Soret/Q-band 吸收與 **Z2** 的藍綠光吸收互補，這類協同夥伴染料不論為溶液態或吸附於 TiO$_2$ 薄膜，皆在 300–700 nm 達到近乎連續的全光譜吸收。由於染料中的紫質及有機單元彼此以非共軛方式連結，因此促使分子內兩單元更緊密且直立地吸附於 TiO$_2$ 表

面,從而提升染料覆蓋度並抑制電荷再結合。同時,**XW60–XW63** 的電化學數據也顯示前沿軌域能階(如 HOMO 與 LUMO)皆與 **XW51** 相近,證明共價連結未改變能階。DSSC 元件效能研究顯示,鏈長過短的 **XW60** 因 **Z2** 無法完整吸附於表面,導致 IPCE 出現缺口而降低效率。至於 **XW61**–**XW63** 元件則顯示出 IPCE 在 400–700 nm 維持約 80%的高平台,有效填補 **XW51** 的吸收缺口。最終效能於 **XW61**–**XW63** 三者皆相近(PCE = 11.6–11.7%;$J_{SC}$ = 20 mA cm$^{-2}$;$V_{OC}$ = 0.76 V),顯示烷氧鏈不論長短皆發揮作用。電化學阻抗譜(electrochemical impedance spectroscopy,EIS)與穩定性測試也顯示 **XW61**–**XW63** 具有較長電子壽命、較低再結合傾向,以及顯著強化的 TiO$_2$ 錨定能力與長期光穩定性。其中,**XW61** 更在近一步添加鵝去氧膽酸(chenodeoxycholic acid,CDCA)作為共吸附劑(co-adsorbent)後,達到 12.4%高光電轉換效率。

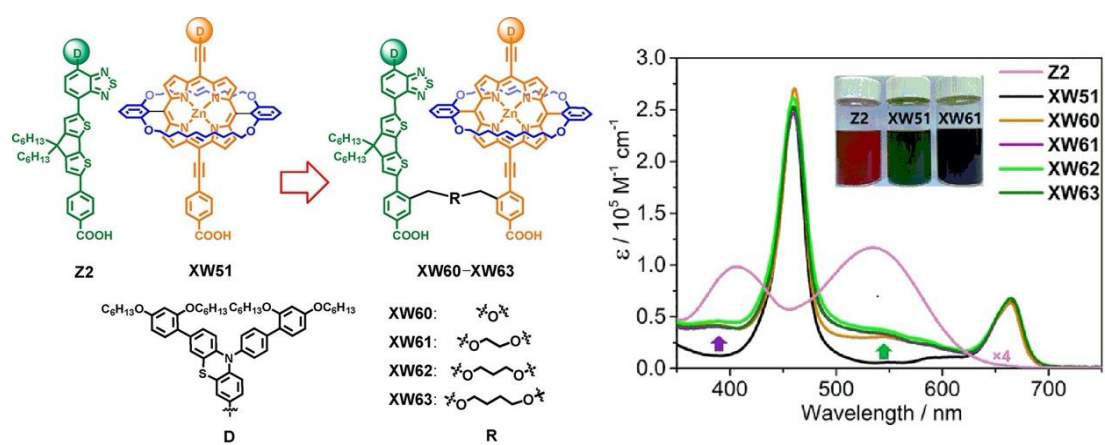

**圖十五、協同夥伴染料的分子結構與吸收光譜圖**[101]。

## 四、全光譜吸收材料的光伏應用與挑戰

在有機太陽能電池(OSCs)中,具全光譜吸收能力的材料有助於提升太陽光譜利用率並增加$J_{SC}$。然而,近年研究指出,單純追求窄能隙以延伸吸光至近紅外光區,並不足以保證低能量損失(energy loss,$E_{loss}$或$\Delta V_{OC}$)與高元件效率。如圖十六(a)所示,儘管窄能隙高分子的能隙分佈在 1–1.5 eV 之間,其 OSC 元件效率普遍落在 10%以下[33]。原因在於此一策略雖可提升光吸收與載子數量,但往往伴隨$V_{OC}$的下降,使整體 PCE 未必隨之提升。如圖十六(b)所示,Menke 等人與 Liu 等人皆指出[102,103],$V_{OC}$與電荷轉移態能量(CT state energy,$E_{CT}$或$E_{PV}$)密切相關,可由以下關係式表示:

$$V_{OC} \approx E_{CT} - V_{loss} \quad (11)$$

其中,$E_{CT}$主要由 donor 的 HOMO 與 acceptor 的 LUMO 能階差所決定:

$$E_{CT} \approx E_{HOMO}^{donor} - E_{LUMO}^{acceptor} \quad (12)$$

若全光譜吸收材料的分子設計僅追求窄能隙，卻導致 donor-acceptor 間能階位移（energetic offsets）過大，反而引入額外的$\Delta V_{OC}$並限制$V_{OC}$。相反地，現代高效率 OSC 的分子設計傾向在確保有效激子解離的前提下，採用較小的能階位移，以兼顧寬光譜吸收與低$V_{loss}$。此外，低能隙與全光譜吸收材料往往伴隨較強的 CT 特徵，若 CT state 過度定域化，將導致非輻射再結合速率（non-radiative recombination rate，$k_{nr}$）上升，進而增加非輻射能量損失（$E_{nr}$），不利於元件效率。Liu 等人指出，$E_{loss}$並非僅由能階排列決定，而與分子結構、界面電子耦合及薄膜形貌密切相關，顯示全光譜材料的設計需同步考量吸光延伸與抑制$E_{nr}$[102]。近年發展的 NFA 材料即透過延伸 π 共軛與精準能階調控，在實現 UV-Vis-NIR 寬光譜吸收的同時，兼顧 CT state 能量與$E_{nr}$控制，使$J_{SC}$與$V_{OC}$得以同時提升，促使單接面 OSC 的光電轉換效率突破 20%[73]。

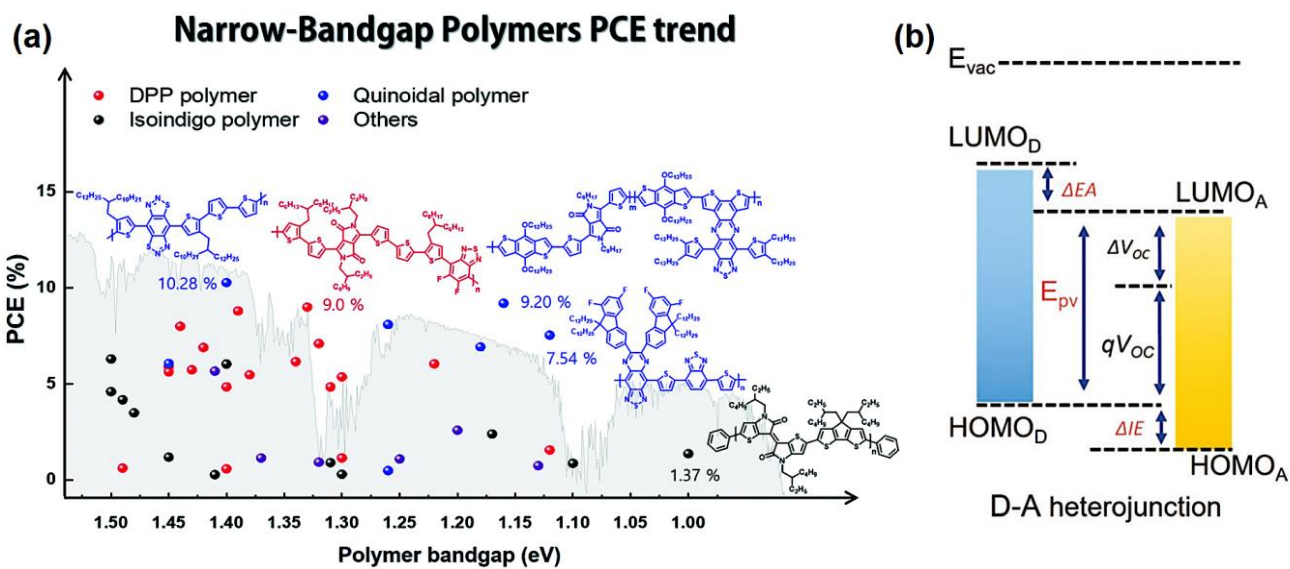

圖十六、(a)窄能隙高分子與其 PCE 之相對關係圖[33]；(b) OSC 元件 D/A 材料的能階關係圖[49]。圖中$\Delta V_{OC}$及$E_{PV}$或稱為$E_{loss}$及$E_{CT}$、$\Delta EA$與$\Delta IE$分別為電子親和力與游離能、$E_{vac}$為真空能階。

另一方面，在染料敏化太陽能電池（DSSCs）系統中，設計全光譜吸收材料不僅取決於染料分子本身，亦高度依賴元件中各個材料間精密的能階匹配[62]。多數情況下，染料分子因採用推-拉型共軛架構而面臨高 HOMO 能階，故全光譜吸收的設計對伴隨而來的 HOMO 能階位移影響有限。然而，由於全光譜吸收染料因窄能隙而不可避免地具備偏低的 LUMO 能階，從而壓縮激發態電子注入 TiO$_2$ 導帶的驅動力（injection free energy，$\Delta G_{inj}$），並大幅抑制注入效率（$\eta_{inj}$）。由前述的公式(4)可知，即便$LHE(\lambda)$得到優化以涵蓋跨波段能量，但過低的$\eta_{inj}$連帶抑制了$\eta_{col}$及$J_{SC}$，從而導致較低的 PCE。由此可知，開發高效率 DSSC 的全光譜吸收染料，並非侷限於染料分子本身，反而面臨各材料間最佳化全盤配置的協同挑戰。其中最關鍵的基礎當從系統層級重新調整半導體、染料及電解質間的能階配置。以下進一步探討兩項策略：

1. **半導體導帶工程**

DSSC 能否採用窄能隙的全光譜染料，維持足夠的$\Delta G_{inj}$是首要瓶頸。因此，從半導體端著手調控導帶位置的導帶工程（conduction band engineering）[104–106]是重要的探討方向。如圖十七(a)所示，採用較低導帶邊緣的半導體材料，如常被研究的 $SnO_2$ 與 $Nb_2O_5$ 等，是具體可行的策略之一[62,107]。另一方面，摻雜（doping）$Nb^{5+}$、$W^{6+}$、$Sn^{4+}$或其他離子[24,107]、進行表面鈍化（passivation）或界面偶極層（dipole layers）等處理[108,109]亦證明可微調 $TiO_2$ 導帶位置，唯過度摻雜將引入缺陷與嚴重再結合問題[62]。

**2. 電解質再生之動力學考量**

由前述公式(8)可知，半導體導帶下移將伴隨$V_{OC}$最大值的壓縮，故電解質系統的協同設計亦無可避免。目前已有大量研究探討不同電解質系統，如氧化還原電位（$E_{redox}$）位於約 +0.4 V vs NHE 的傳統$I^-/I_3^-$電解質對、以及$E_{redox}$可大幅調控的 $Co^{2+}/Co^{3+}$及 $Cu^+/Cu^{2+}$氧化還原穿梭體（redox shuttle）系統[9,110]，如圖十七(b)所示。其能階及氧化還原特性與$\eta_{reg}$息息相關。有趣的是，電解質效能並非僅由$E_{redox}$數值高低決定，原因在於染料再生速率可能落於前述馬可斯電子轉移理論的一般區或反轉區（normal/inverted region），並進一步影響電解質的自交換速率（self-exchange rate）與質傳行為。因此，在進行半導體導帶工程的同時，也綜合考量再生驅動力（regeneration free energy，$\Delta G_{reg}$）、重組能（reorganization energy，$\lambda$）、電子耦合強度（electronic coupling，$V_{DA}$），藉以優化元件材料間的電子傳遞速率（$k_{et}$），才是同時提升$V_{OC}$與$J_{SC}$的關鍵。

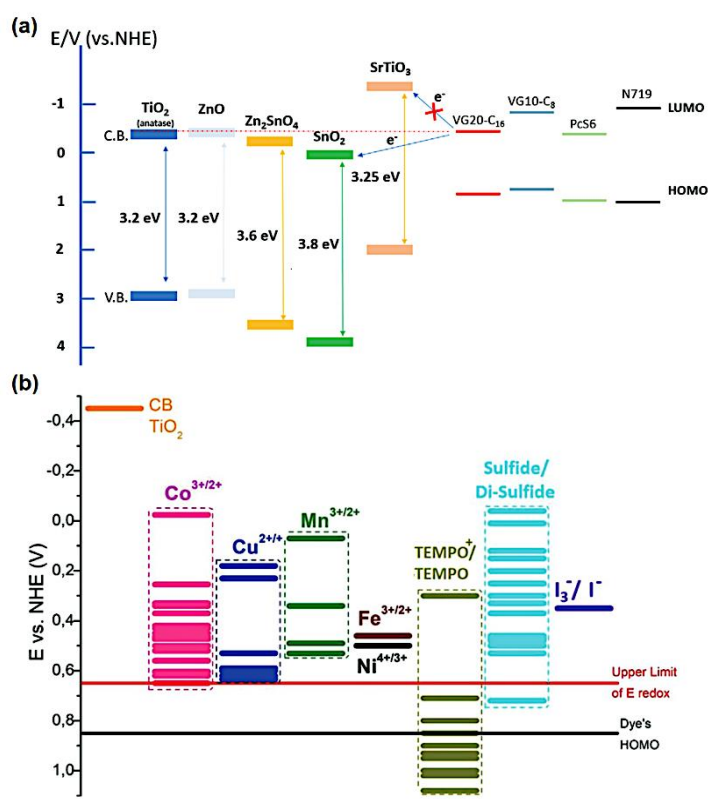

圖十七、(a)氧化物半導體價帶、導帶與低能隙染料之 LUMO 能階示意圖[62]。

## 五、結語

綜合而言，全光譜吸收材料在光伏應用中的挑戰，並非僅源自吸光能力本身，而是深植於能階排列、界面電子耦合與電荷動力學之間的精密平衡。因此，在材料設計過程跳脫單一分子設計，並轉而從元件系統層級協同調控吸收、注入、再生與能量損失等機制，方能真正釋放全光譜吸收材料於高效率光伏元件中的潛力。

## 參考文獻

[1]  G. D. Scholes, G. R. Fleming, A. Olaya-Castro, R. Van Grondelle, "Lessons from nature about solar light harvesting" *Nature Chem* **2011**, *3*, 763–774.

[2]  P. Fromme, H. Yu, Y. S. DeRuyter, C. Jolley, D. K. Chauhan, A. Melkozernov, I. Grotjohann, "Structure of photosystems I and II" *Comptes Rendus. Chimie* **2005**, *9*, 188–200.

[3]  D. Saccone, S. Galliano, N. Barbero, P. Quagliotto, G. Viscardi, C. Barolo, "Polymethine Dyes in Hybrid Photovoltaics: Structure–Properties Relationships" *Eur J Org Chem* **2016**, *2016*, 2244–2259.

[4]  A. Mishra, P. Baeuerle, "Small Molecule Organic Semiconductors on the Move: Promises for Future Solar Energy Technology" *Angew Chem Int Ed* **2012**, *51*, 2020–2067.

[5]  Z. Li, Q. Li, C. Li, Y. Xie, "Panchromatic porphyrin-based dye-sensitized solar cells: from cosensitization to concerted companion dye approaches" *Mater. Chem. Front.* **2024**, *8*, 652–680.

[6]  B. Kippelen, J.-L. Bredas, "Organic photovoltaics" *Energy Environ. Sci.* **2009**, *2*, 251.

[7]  M. Grätzel, "Solar Energy Conversion by Dye-Sensitized Photovoltaic Cells" *Inorg. Chem.* **2005**, *44*, 6841–6851.

[8]  G. Boschloo, A. Hagfeldt, "Characteristics of the Iodide/Triiodide Redox Mediator in Dye-Sensitized Solar Cells" *Acc. Chem. Res.* **2009**, *42*, 1819–1826.

[9]  T. Higashino, H. Imahori, "Emergence of Copper(I/II) Complexes as Third-Generation Redox Shuttles for Dye-Sensitized Solar Cells" *ACS Energy Lett.* **2022**, *7*, 1926–1938.

[10] M. Wang, C. Graetzel, S. M. Zakeeruddin, M. Grätzel, "Recent developments in redox electrolytes for dye-sensitized solar cells" *Energy Environ. Sci.* **2012**, *5*, 9394–9405.

[11] V. Gondane, P. Bhargava, "Tuning flat band potential of TiO 2 using an electrolyte additive to enhance open circuit voltage and minimize current loss in dye sensitized solar cells" *Electrochimica Acta* **2016**, *209*, 293–298.

[12] G. Boschloo, L. Häggman, A. Hagfeldt, "Quantification of the Effect of 4-*tert*-Butylpyridine Addition to I$^-$/I$_3^-$ Redox Electrolytes in Dye-Sensitized Nanostructured TiO$_2$ Solar Cells" *J. Phys. Chem. B* **2006**, *110*, 13144–13150.

[13] S. Nakade, T. Kanzaki, W. Kubo, T. Kitamura, Y. Wada, S. Yanagida, "Role of Electrolytes on Charge Recombination in Dye-Sensitized TiO$_2$ Solar Cell (1): The Case of Solar Cells Using the I$^-$/I$_3^-$ Redox Couple" *J. Phys. Chem. B* **2005**, *109*, 3480–3487.

[14] Y. H. Lee, R. K. Chitumalla, B. Y. Jang, J. Jang, S. Thogiti, J. H. Kim, "Alkyl chain length dependence of the charge-transfer, recombination and electron diffusion length on the photovoltaic performance in double donor-acceptor-based organic dyes for dye sensitized solar cells" *Dyes and Pigments* **2016**, *133*, 161–172.

[15] M. Suzuka, N. Hayashi, T. Sekiguchi, K. Sumioka, M. Takata, N. Hayo, H. Ikeda, K. Oyaizu, H. Nishide, "A Quasi-Solid State DSSC with 10.1% Efficiency through Molecular Design of the Charge-Separation and -Transport" *Sci Rep* **2016**, *6*, 28022.

[16] J. E. Kroeze, N. Hirata, S. Koops, Md. K. Nazeeruddin, L. Schmidt-Mende, M. Grätzel, J. R. Durrant, "Alkyl Chain Barriers for Kinetic Optimization in Dye-Sensitized Solar Cells" *J. Am. Chem. Soc.* **2006**, *128*, 16376–16383.

[17] S. Yoon, E.-Y. Shin, N.-K. Cho, S. Park, H. Y. Woo, H. J. Son, "Progress in morphology control from fullerene to nonfullerene acceptors for scalable high-performance organic photovoltaics" *J. Mater. Chem. A* **2021**, *9*, 24729–24758.

[18] S. Zhang, X. Yang, Y. Numata, L. Han, "Highly efficient dye-sensitized solar cells: progress and future challenges" *Energy Environ. Sci.* **2013**, *6*, 1443.

[19] M. Aftabuzzaman, S. Sarker, C. Lu, H. K. Kim, "In-depth understanding of the energy loss and efficiency limit of dye-sensitized solar cells under outdoor and indoor conditions" *J. Mater. Chem. A* **2021**, *9*, 24830–24848.

[20] H. Bronstein, C. B. Nielsen, B. C. Schroeder, I. McCulloch, "The role of chemical design in the performance of organic semiconductors" *Nat Rev Chem* **2020**, *4*, 66–77.

[21] O. Ostroverkhova, "Organic Optoelectronic Materials: Mechanisms and Applications" *Chem. Rev.* **2016**, *116*, 13279–13412.

[22] R. A. Marcus, "On the Theory of Oxidation-Reduction Reactions Involving Electron Transfer. I" *The Journal of Chemical Physics* **1956**, *24*, 966–978.

[23] A. Atxabal, T. Arnold, S. Parui, S. Hutsch, E. Zuccatti, R. Llopis, M. Cinchetti, F. Casanova, F. Ortmann, L. E. Hueso, "Tuning the charge flow between Marcus regimes in an organic thin-film device" *Nat Commun* **2019**, *10*, 2089.

[24] K. Sharma, V. Sharma, S. S. Sharma, "Dye-Sensitized Solar Cells: Fundamentals and Current Status" *Nanoscale Res Lett* **2018**, *13*, 381.

[25] A. Kahn, "Fermi level, work function and vacuum level" *Mater. Horiz.* **2016**, *3*, 7–10.

[26] L. Li, T. Li, Y. Lin, "Organic photovoltaic materials with near-infrared II-region response" *Giant* **2024**, *20*, 100334.

[27] J. A. Pincock, A. W. Speed, "The aromatic character and resonance stabilization energies of substituted cyclopentadienyl and indenyl cations" *Can. J. Chem.* **2005**, *83*, 1287–1298.

[28] S. Olsen, R. H. McKenzie, "Bond alternation, polarizability, and resonance detuning in methine dyes" *The Journal of Chemical Physics* **2011**, *134*, 114520.

[29] N. A. Murugan, J. Kongsted, Z. Rinkevicius, H. Ågren, "Breakdown of the first hyperpolarizability/bond-length alternation parameter relationship" *Proc. Natl. Acad. Sci. U.S.A.* **2010**, *107*, 16453–16458.

[30] S. R. Marder, C. B. Gorman, B. G. Tiemann, J. W. Perry, G. Bourhill, K. Mansour, "Relation Between Bond-Length Alternation and Second Electronic Hyperpolarizability of Conjugated Organic Molecules" *Science* **1993**, *261*, 186–189.

[31] B. He, D. Zherebetskyy, H. Wang, M. A. Kolaczkowski, L. M. Klivansky, T. Tan, L. Wang, Y. Liu, "Rational tuning of high-energy visible light absorption for panchromatic small molecules by a two-dimensional conjugation approach" *Chem. Sci.* **2016**, *7*, 3857–3861.

[32] Z. Chen, W. Li, M. A. Sabuj, Y. Li, W. Zhu, M. Zeng, C. S. Sarap, M. M. Huda, X. Qiao, X. Peng, D. Ma, Y. Ma, N. Rai, F. Huang, "Evolution of the electronic structure in open-shell donor-acceptor organic semiconductors" *Nat Commun* **2021**, *12*, 5889.

[33] D.-H. Lim, J.-W. Ha, H. Choi, S. C. Yoon, B. R. Lee, S.-J. Ko, "Recent progress of ultra-narrow-bandgap polymer donors for NIR-absorbing organic solar cells" *Nanoscale Adv.* **2021**, *3*, 4306–4320.

[34] R. Martin, S. L. Buchwald, "Palladium-Catalyzed Suzuki−Miyaura Cross-Coupling Reactions Employing Dialkylbiaryl Phosphine Ligands" *Acc. Chem. Res.* **2008**, *41*, 1461–1473.

[35] C. Cordovilla, C. Bartolomé, J. M. Martínez-Ilarduya, P. Espinet, "The Stille Reaction, 38 Years Later" *ACS Catal.* **2015**, *5*, 3040–3053.

[36] R. Chinchilla, C. Nájera, "The Sonogashira Reaction: A Booming Methodology in Synthetic Organic Chemistry" *Chem. Rev.* **2007**, *107*, 874–922.

[37] R. Dorel, C. P. Grugel, A. M. Haydl, "The Buchwald–Hartwig Amination After 25 Years" *Angew Chem Int Ed* **2019**, *58*, 17118–17129.

[38] K. Van Beurden, S. De Koning, D. Molendijk, J. Van Schijndel, "The Knoevenagel reaction: a review of the unfinished treasure map to forming carbon–carbon bonds" *Green Chemistry Letters and Reviews* **2020**, *13*, 349–364.

[39] X. Zhong, S. Liu, W. You, "A general and mild synthetic method for fused-ring electronic acceptors" *Sci. Adv.* **2024**, *10*, eadp8150.

[40] M. Grzybowski, K. Skonieczny, H. Butenschön, D. T. Gryko, "Comparison of Oxidative Aromatic Coupling and the Scholl Reaction" *Angew Chem Int Ed* **2013**, *52*, 9900–9930.

[41] H. Bohra, M. Wang, "Direct C–H arylation: a 'Greener' approach towards facile synthesis of organic semiconducting molecules and polymers" *J. Mater. Chem. A* **2017**, *5*, 11550–11571.

[42] P. Brogdon, H. Cheema, J. H. Delcamp, "Near-Infrared-Absorbing Metal-Free Organic, Porphyrin, and Phthalocyanine Sensitizers for Panchromatic Dye-Sensitized Solar Cells" *ChemSusChem* **2018**, *11*, 86–103.

[43] Q. Huaulmé, C. Aumaitre, O. V. Kontkanen, D. Beljonne, A. Sutter, G. Ulrich, R. Demadrille, N. Leclerc, "Functional panchromatic BODIPY dyes with near-infrared absorption: design, synthesis, characterization and use in dye-sensitized solar cells" *Beilstein J. Org. Chem.* **2019**, *15*, 1758–1768.

[44] A. Tang, C. Zhan, J. Yao, E. Zhou, "Design of Diketopyrrolopyrrole (DPP)-Based Small Molecules for Organic-Solar-Cell Applications" *Advanced Materials* **2017**, *29*, 1600013.

[45] Y. Wu, W. Zhu, "Organic sensitizers from D–π–A to D–A–π–A: effect of the internal electron-withdrawing units on molecular absorption, energy levels and photovoltaic performances" *Chem. Soc. Rev.* **2013**, *42*, 2039–2058.

[46] Z. Wang, S. Zhu, T. Li, W. Liang, J. Zhou, H. Hu, "Fluorinated Benzothiadiazole-Based Polymers for Organic Solar Cells: Progress and Prospects" *ACS Mater. Au* **2025**, *5*, 57–71.

[47] J. Feng, W. Jiang, Z. Wang, "Synthesis and Application of Rylene Imide Dyes as Organic Semiconducting Materials" *Chemistry An Asian Journal* **2018**, *13*, 20–30.

[48] B. O'Regan, M. Graetzel, "A low-cost, high-efficiency solar cell based on dye-sensitized colloidal TiO2 films" *Nature* **1991**, *353*, 737–740.

[49] J. Bertrandie, J. Han, C. S. P. De Castro, E. Yengel, J. Gorenflot, T. Anthopoulos, F. Laquai, A. Sharma, D. Baran, "The Energy Level Conundrum of Organic Semiconductors in Solar Cells" *Advanced Materials* **2022**, *34*, 2202575.

[50] H.-H. Chou, C.-H. Yang, J. T. Lin, C.-P. Hsu, "First-Principle Determination of Electronic Coupling and Prediction of Charge Recombination Rates in Dye-Sensitized Solar Cells" *J. Phys. Chem. C* **2017**, *121*, 983–992.

[51] L. Perdigón-Toro, H. Zhang, A. Markina, J. Yuan, S. M. Hosseini, C. M. Wolff, G. Zuo, M. Stolterfoht, Y. Zou, F. Gao, D. Andrienko, S. Shoaee, D. Neher, "Barrierless Free Charge Generation in the High-Performance PM6:Y6 Bulk Heterojunction Non-Fullerene Solar Cell" *Advanced Materials* **2020**, *32*, 1906763.

[52] S. M. Pratik, G. Kupgan, J.-L. Brédas, V. Coropceanu, "Analysis of the charge generation and recombination processes in the PM6:Y6 organic solar cell" *Energy Environ. Sci.* **2025**, *18*, 841–852.

[53] K. Neumann, C. Schwarz, A. Köhler, M. Thelakkat, "Influence of the Excited-State Charge-Transfer Character on the Exciton Dissociation in Donor–Acceptor Copolymers" *J. Phys. Chem. C* **2014**, *118*, 27–36.

[54] Y. Wang, J. Ren, Z. Shuai, "Minimizing non-radiative decay in molecular aggregates through control of excitonic coupling" *Nat Commun* **2023**, *14*, 5056.

[55] J. Watson, T. J. Santaloci, H. Cheema, R. C. Fortenberry, J. H. Delcamp, "Full Visible Spectrum Panchromatic Triple Donor Dye for Dye-Sensitized Solar Cells" *J. Phys. Chem. C* **2020**, *124*, 25211–25220.

[56] Y. Ooyama, M. Kanda, T. EnoKi, Y. Adachi, J. Ohshita, "Synthesis, optical and electrochemical properties, and photovoltaic performance of a panchromatic and near-infrared (D)$_2$ –π–A type BODIPY dye with pyridyl group or cyanoacrylic acid" *RSC Adv.* **2017**, *7*, 13072–13081.

[57] L.-L. Li, E. W.-G. Diau, "Porphyrin-sensitized solar cells" *Chem. Soc. Rev.* **2013**, *42*, 291–304.

[58] A. Roy, N. C. M. Magdaong, H. Jing, J. Rong, J. R. Diers, H. S. Kang, D. M. Niedzwiedzki, M. Taniguchi, C. Kirmaier, J. S. Lindsey, D. F. Bocian, D. Holten, "Balancing Panchromatic Absorption and Multistep Charge Separation in a Compact Molecular Architecture" *J. Phys. Chem. A* **2022**, *126*, 9353–9365.

[59] G. Hu, H. S. Kang, A. K. Mandal, A. Roy, C. Kirmaier, D. F. Bocian, D. Holten, J. S. Lindsey, "Synthesis of arrays containing porphyrin, chlorin, and perylene-imide constituents for panchromatic light-harvesting and charge separation" *RSC Adv.* **2018**, *8*, 23854–23874.

[60] P. Cheng, X. Zhao, X. Zhan, "Perylene Diimide-Based Oligomers and Polymers for Organic Optoelectronics" *Acc. Mater. Res.* **2022**, *3*, 309–318.

[61] A. Schulz, R. Fröhlich, A. Jayachandran, F. Schneider, M. Stolte, T. Brixner, F. Würthner, "Panchromatic light-harvesting antenna by supramolecular exciton band engineering for heteromeric dye foldamer" *Chem* **2024**, *10*, 2887–2900.

[62] F. Grifoni, M. Bonomo, W. Naim, N. Barbero, T. Alnasser, I. Dzeba, M. Giordano, A. Tsaturyan, M. Urbani, T. Torres, C. Barolo, F. Sauvage, "Toward Sustainable, Colorless, and Transparent Photovoltaics: State of the Art and Perspectives for the Development of Selective Near-Infrared Dye-Sensitized Solar Cells" *Adv. Energy Mater.* **2021**, *11*, 2101598.

[63] E. N. Bifari, P. Almeida, R. M. El-Shishtawy, "Advancing panchromatic effect for efficient sensitization of cyanine and hemicyanine-based dye-sensitized solar cells" *Materials Today Energy* **2023**, *36*, 101337.

[64] W. Naim, V. Novelli, I. Nikolinakos, N. Barbero, I. Dzeba, F. Grifoni, Y. Ren, T. Alnasser, A. Velardo, R. Borrelli, S. Haacke, S. M. Zakeeruddin, M. Graetzel, C. Barolo, F. Sauvage, "Transparent and Colorless Dye-Sensitized Solar Cells Exceeding 75% Average Visible Transmittance" *JACS Au* **2021**, *1*, 409–426.

[65] I. Osaka, M. Shimawaki, H. Mori, I. Doi, E. Miyazaki, T. Koganezawa, K. Takimiya, "Synthesis, Characterization, and Transistor and Solar Cell Applications of a Naphthobisthiadiazole-Based Semiconducting Polymer" *J. Am. Chem. Soc.* **2012**, *134*, 3498–3507.

[66] I. Osaka, K. Takimiya, "Naphthobischalcogenadiazole Conjugated Polymers: Emerging Materials for Organic Electronics" *Advanced Materials* **2017**, *29*, 1605218.

[67] J. W. Jung, F. Liu, T. P. Russell, W. H. Jo, "A high mobility conjugated polymer based on dithienothiophene and diketopyrrolopyrrole for organic photovoltaics" *Energy Environ. Sci.* **2012**, *5*, 6857–6861.

[68] J. Yuan, Y. Zhang, L. Zhou, G. Zhang, H.-L. Yip, T.-K. Lau, X. Lu, C. Zhu, H. Peng, P. A. Johnson, M. Leclerc, Y. Cao, J. Ulanski, Y. Li, Y. Zou, "Single-Junction Organic Solar Cell with over 15% Efficiency Using Fused-Ring Acceptor with Electron-Deficient Core" *Joule* **2019**, *3*, 1140–1151.

[69] P. Cheng, G. Li, X. Zhan, Y. Yang, "Next-generation organic photovoltaics based on non-fullerene acceptors" *Nature Photon* **2018**, *12*, 131–142.

[70] C. Yan, S. Barlow, Z. Wang, H. Yan, A. K.-Y. Jen, S. R. Marder, X. Zhan, "Non-fullerene acceptors for organic solar cells" *Nat Rev Mater* **2018**, *3*, 18003.

[71] J. Hou, O. Inganäs, R. H. Friend, F. Gao, "Organic solar cells based on non-fullerene acceptors" *Nat. Mater.* **2018**, *17*, 119–128.

[72] R. Yu, G. Wu, Z. Tan, "Realization of high performance for PM6:Y6 based organic photovoltaic cells" *Journal of Energy Chemistry* **2021**, *61*, 29–46.

[73] H. Chen, Y. Huang, R. Zhang, H. Mou, J. Ding, J. Zhou, Z. Wang, H. Li, W. Chen, J. Zhu, Q. Cheng, H. Gu, X. Wu, T. Zhang, Y. Wang, H. Zhu, Z. Xie, F. Gao, Y. Li, Y. Li, "Organic solar cells with 20.82% efficiency and high tolerance of active layer thickness through crystallization sequence manipulation" *Nat. Mater.* **2025**, *24*, 444–453.

[74] Q. Li, Y. Zhang, Z. Xie, Y. Zhen, W. Hu, H. Dong, "Polycyclic aromatic hydrocarbon-based organic semiconductors: ring-closing synthesis and optoelectronic properties" *J. Mater. Chem. C* **2022**, *10*, 2411–2430.

[75] P. Murugan, E. Ravindran, V. Sangeetha, S.-Y. Liu, J. W. Jung, "Perylene-diimide for organic solar cells: current scenario and prospects in molecular geometric, functionalization, and optoelectronic properties" *J. Mater. Chem. A* **2023**, *11*, 26393–26425.

[76] N. Liang, D. Meng, Z. Wang, "Giant Rylene Imide-Based Electron Acceptors for Organic Photovoltaics" *Acc. Chem. Res.* **2021**, *54*, 961–975.

[77] A. Bhardwaj, Ch. Mudasar Hussain, P. Dewangan, P. Mukhopadhyay, "Naphthalene diimide–Annulated Heterocyclic Acenes: Synthesis, Electrochemical and Semiconductor Properties and their Multifaceted Applications" *Chemistry A European J* **2024**, *30*, e202400208.

[78] C.-L. Mai, W.-K. Huang, H.-P. Lu, C.-W. Lee, C.-L. Chiu, Y.-R. Liang, E. W.-G. Diau, C.-Y. Yeh, "Synthesis and characterization of diporphyrin sensitizers for dye-sensitized solar cells" *Chem. Commun.* **2010**, *46*, 809–811.

[79] Q. Zhao, H. Lai, H. Chen, H. Li, F. He, "H- and J-aggregation inspiring efficient solar conversion" *J. Mater. Chem. A* **2021**, *9*, 1119–1126.

[80] M. S. Iqbal, H. Ahmad, C. Yu, H. Huang, B. Guo, "Near-infrared BODIPY-based theranostic agents for photothermal and combinatory therapy of cancers, where do we stand in year 2025" *Coordination Chemistry Reviews* **2025**, *541*, 216781.

[81] M. Más-Montoya, R. A. J. Janssen, "The Effect of H- and J-Aggregation on the Photophysical and Photovoltaic Properties of Small Thiophene-Pyridine-DPP Molecules for Bulk-Heterojunction Solar Cells" *Adv. Funct. Mater.* **2017**, *27*, 1605779.

[82] C. Liao, K. Zeng, H. Wu, Q. Zeng, H. Tang, L. Wang, H. Meier, Y. Xie, D. Cao, "Conjugating pillararene dye in dye-sensitized solar cells" *Cell Reports Physical Science* **2021**, *2*, 100326.

[83] S. V. Bhosale, M. Al Kobaisi, R. W. Jadhav, P. P. Morajkar, L. A. Jones, S. George, "Naphthalene diimides: perspectives and promise" *Chem. Soc. Rev.* **2021**, *50*, 9845–9998.

[84] M. R. Biradar, S. V. Bhosale, P. P. Morajakar, S. V. Bhosale, "A review on energy storage devices based on rylene imide dyes: Synthesis, applications and challenges" *Fuel* **2022**, *310*, 122487.

[85] H. Zhang, C. Liu, Z. Wang, X. Liu, Z. Han, X. Zhang, Y. Li, Q. Zhao, G. He, "Synergistic ionic modification strategy enhances the stability of naphthalene diimide zwitterions for cost-effective aqueous organic redox flow batteries" *National Science Review* **2025**, *12*, nwaf123.

[86] S. Takenaka, "Application of naphthalene diimide in biotechnology" *Polym J* **2021**, *53*, 415–427.

[87] S. Suraru, F. Wuerthner, "Strategies for the Synthesis of Functional Naphthalene Diimides" *Angew Chem Int Ed* **2014**, *53*, 7428–7448.

[88] T. Roy, I. Debnath, K. Mahata, "Synthesis, optical properties and cation mediated tuning of the reduction potentials of core-annulated naphthalene diimide derivatives" *Org. Chem. Front.* **2022**, *9*, 3255–3261.

[89] K. A. Wells, J. R. Palmer, J. E. Yarnell, S. Garakyaraghi, B. C. Pemberton, J. M. Favale, M. K. Valchar, A. Chakraborty, F. N. Castellano, "Understanding the influence of geometric and

electronic structure on the excited state dynamical and photoredox properties of perinone chromophores" *Phys. Chem. Chem. Phys.* **2021**, *23*, 24200–24210.

[90] M. Mamada, C. Pérez-Bolívar, D. Kumaki, N. A. Esipenko, S. Tokito, P. Anzenbacher, "Benzimidazole Derivatives: Synthesis, Physical Properties, and n-Type Semiconducting Properties" *Chemistry A European J* **2014**, *20*, 11835–11846.

[91] W. Jiang, Y. Li, Z. Wang, "Tailor-Made Rylene Arrays for High Performance n-Channel Semiconductors" *Acc. Chem. Res.* **2014**, *47*, 3135–3147.

[92] S.-L. Suraru, C. Burschka, F. Wuerthner, "Diindole-Annulated Naphthalene Diimides: Synthesis and Optical and Electronic Properties of *Syn* - and *Anti* -Isomers" *J. Org. Chem.* **2014**, *79*, 128–139.

[93] Y. Chi, K. Wu, T. Wei, "Ruthenium and Osmium Complexes That Bear Functional Azolate Chelates for Dye-Sensitized Solar Cells" *Chemistry An Asian Journal* **2015**, *10*, 1098–1115.

[94] R. Juwita, J.-M. Liao, C.-Y. Chen, H.-H. G. Tsai, "Enhancing Near-Infrared Absorption in Terpyridyl Ru/Os Complexes with Ancillary Ligands to Activate Spin-Forbidden Transitions in Dye-Sensitized Solar Cells: A TDDFT Investigation" *J. Phys. Chem. A* **2024**, *128*, 880–894.

[95] S.-H. Liu, H. Fu, Y.-M. Cheng, K.-L. Wu, S.-T. Ho, Y. Chi, P.-T. Chou, "Theoretical Study of N749 Dyes Anchoring on the $(TiO_2)_{28}$ Surface in DSSCs and Their Electronic Absorption Properties" *J. Phys. Chem. C* **2012**, *116*, 16338–16345.

[96] G. De La Torre, G. Bottari, T. Torres, "Phthalocyanines and Subphthalocyanines: Perfect Partners for Fullerenes and Carbon Nanotubes in Molecular Photovoltaics" *Advanced Energy Materials* **2017**, *7*, 1601700.

[97] R. Juwita, J.-Y. Lin, S.-J. Lin, Y.-C. Liu, T.-Y. Wu, Y.-M. Feng, C.-Y. Chen, H.-H. Gavin Tsai, C.-G. Wu, "Osmium sensitizer with enhanced spin–orbit coupling for panchromatic dye-sensitized solar cells" *J. Mater. Chem. A* **2020**, *8*, 12361–12369.

[98] J. M. Cole, G. Pepe, O. K. Al Bahri, C. B. Cooper, "Cosensitization in Dye-Sensitized Solar Cells" *Chem. Rev.* **2019**, *119*, 7279–7327.

[99] A. Yella, H.-W. Lee, H. N. Tsao, C. Yi, A. K. Chandiran, M. K. Nazeeruddin, E. W.-G. Diau, C.-Y. Yeh, S. M. Zakeeruddin, M. Gratzel, "Porphyrin-Sensitized Solar Cells with Cobalt (II/III)-Based Redox Electrolyte Exceed 12 Percent Efficiency" *Science* **2011**, *334*, 629–634.

[100] C. B. Cooper, E. J. Beard, Á. Vázquez-Mayagoitia, L. Stan, G. B. G. Stenning, D. W. Nye, J. A. Vigil, T. Tomar, J. Jia, G. B. Bodedla, S. Chen, L. Gallego, S. Franco, A. Carella, K. R. J. Thomas, S. Xue, X. Zhu, J. M. Cole, "Design-to-Device Approach Affords Panchromatic Co-Sensitized Solar Cells" *Advanced Energy Materials* **2019**, *9*, 1802820.

[101] K. Zeng, Y. Chen, W.-H. Zhu, H. Tian, Y. Xie, "Efficient Solar Cells Based on Concerted Companion Dyes Containing Two Complementary Components: An Alternative Approach for Cosensitization" *J. Am. Chem. Soc.* **2020**, *142*, 5154–5161.

[102] Z. Liu, J. Xin, X. Liu, Y. Zhang, M. Duan, Y. Zhang, Z. Xu, J. Yan, Q. Liang, J. Liu, "Minimizing energy losses in organic solar cells: mechanistic insights, material design, and morphological control" *EES Sol.* **2025**, 10.1039.D5EL00138B.

[103] S. M. Menke, N. A. Ran, G. C. Bazan, R. H. Friend, "Understanding Energy Loss in Organic Solar Cells: Toward a New Efficiency Regime" *Joule* **2018**, *2*, 25–35.

[104] P. S. Saud, A. Bist, A. A. Kim, A. Yousef, A. Abutaleb, M. Park, S.-J. Park, B. Pant, "Dye-sensitized solar cells: Fundamentals, recent progress, and Optoelectrical properties improvement strategies" *Optical Materials* **2024**, *150*, 115242.

[105] M. Dürr, S. Rosselli, A. Yasuda, G. Nelles, "Band-Gap Engineering of Metal Oxides for Dye-Sensitized Solar Cells" *J. Phys. Chem. B* **2006**, *110*, 21899–21902.

[106] R. Jose, V. Thavasi, S. Ramakrishna, "Metal Oxides for Dye-Sensitized Solar Cells" *Journal of the American Ceramic Society* **2009**, *92*, 289–301.

[107] A. Agrawal, S. A. Siddiqui, A. Soni, G. D. Sharma, "Advancements, frontiers and analysis of metal oxide semiconductor, dye, electrolyte and counter electrode of dye sensitized solar cell" *Solar Energy* **2022**, *233*, 378–407.

[108] A. B. Muñoz-García, I. Benesperi, G. Boschloo, J. J. Concepcion, J. H. Delcamp, E. A. Gibson, G. J. Meyer, M. Pavone, H. Pettersson, A. Hagfeldt, M. Freitag, "Dye-sensitized solar cells strike back" *Chem. Soc. Rev.* **2021**, *50*, 12450–12550.

[109] N. Kopidakis, N. R. Neale, A. J. Frank, "Effect of an Adsorbent on Recombination and Band-Edge Movement in Dye-Sensitized TiO$_2$ Solar Cells: Evidence for Surface Passivation" *J. Phys. Chem. B* **2006**, *110*, 12485–12489.

[110] Masud, H. K. Kim, "Redox Shuttle-Based Electrolytes for Dye-Sensitized Solar Cells: Comprehensive Guidance, Recent Progress, and Future Perspective" *ACS Omega* **2023**, *8*, 6139–6163.